\documentclass{aa}    % if the references are not structured
\usepackage{graphicx}
\usepackage{txfonts}
\usepackage[T1]{fontenc}

\DeclareUnicodeCharacter{2061}{}
\DeclareUnicodeCharacter{2062}{}

\usepackage{color}
\usepackage[dvipsnames]{xcolor}

\usepackage[pdfencoding=auto, psdextra, unicode]{hyperref}

\definecolor{ballblue}{rgb}{0.13, 0.67, 0.8}

\hypersetup{colorlinks=true,linkcolor=NavyBlue,filecolor=NavyBlue,urlcolor=NavyBlue,citecolor=NavyBlue}

\usepackage{graphicx}	
\usepackage{amsmath}

\begin{document} 

   \title{The relationship between warm and hot has-phase metallicity in massive elliptical galaxies and the influence of AGN feedback}
   \subtitle{}

   \author{Valeria Olivares
          \inst{1,}\inst{2}
          \and Yuanyuan Su\inst{3}
         \and Pasquale Temi\inst{4}
          \and Ryan Eskenasy\inst{3}
           \and Helen Russell\inst{5}
           \and Massimo Gaspari\inst{6}
            \and Philippe Salome\inst{7}
            \and Francoise Combes\inst{7}    
            \and Ming Sun\inst{8}
            \and Ezequiel Treister\inst{9}
            \and Kevin Fogarty\inst{4}
            \and Ana Jimenez-Gallardo\inst{10}
            \and Patricio Lagos\inst{11}
          }

   \institute{Departamento de F\'isica, Universidad de Santiago de Chile, Av. Victor Jara 3659, Santiago 9170124, Chile \email{valeria.olivares@usach.cl}
        \and 
        Center for Interdisciplinary Research in Astrophysics and Space Exploration (CIRAS), Universidad de Santiago de Chile, Santiago 9170124, Chile 
         \and 
        Department of Physics and Astronomy, University of Kentucky, 505 Rose Street, Lexington, KY 40506, USA 
        \and
        Astrophysics Branch, NASA-Ames Research Center, MS 245-6, Moffett Field, CA 94035, USA 
        \and 
        School of Physics \& Astronomy, University of Nottingham, University Park, Nottingham NG7 2RD, UK 
        \and 
        Department of Physics, Informatics \& Mathematics, University of Modena \& Reggio Emilia, 41125 Modena, Italy 
        \and
        LERMA, Observatoire de Paris, Collége de France, PSL University, CNRS, Sorbonne University, Paris, France 
        \and
        Department of Physics and Astronomy, University of Alabama in Huntsville, Huntsville, AL 35899, USA 
        \and
        Instituto de Alta Investigaci{\'{o}}n, Universidad de Tarapac{\'{a}}, Casilla 7D, Arica, Chile
        \and
        European Southern Observatory, Alonso de Córdova 3107, Vitacura, Región Metropolitana, Chile 
        \and
        Institute of Astrophysics, Facultad de Ciencias Exactas, Universidad Andrés Bello, Sede Concepción, Talcahuano, Chile 
        }

 %  \date{Received 07/04/2025; accepted 07/11/2025}

 \abstract{
{\tt Context.} 
Warm ionized gas is ubiquitous at the centers of X-ray bright elliptical galaxies. While it is believed to play a key role in the feeding and feedback processes of supermassive black holes, its origins remain under debate. Existing studies have primarily focused on the morphology and kinematics of warm ionized gas.

{\tt Aims.} 
This work aims to provide a new perspective on warm ($\sim$10,000~K) ionized gas and its connection to X-ray-emitting hot gas ($>$10$^{6}$K) by measuring and comparing their metallicities.

{\tt Methods.} 
We conducted a joint analysis of 13 massive elliptical galaxies using MUSE/VLT and \textit{Chandra} observations. Emission-line ratios, including $\rm [OIII]/H\beta$, $\rm [NII]/H\alpha$, and $\rm [SII]/H\alpha$, were measured for the warm ionized gas using MUSE observations. These ratios were used to infer the ionization mechanisms and derive metallicities of the warm ionized gas using HII, LINER/LIER, and AGN calibrations. We also computed the warm phase metallicity using X-ray/EUV, and pAGB stars models. For two sources at higher redshift, direct $T_e$ method was also used to measure warm gas metallicities. The metallicity of the hot gas was measured using \textit{Chandra} X-ray observations.

{\tt Results.} 
Our observations reveal that most sources exhibit composite ionization, with contributions from both star formation and LINER-like emission. The four sources with the lowest star formation rates in our sample—Centaurus, M87, M84, Abell\,496—are dominated by LINER emission. A positive linear correlation was found between the gas-phase metallicities of the warm and hot phases, ranging from 0.3 to 1.5 $Z_{\odot}$. In some sources the warm gas metallicity shows a central drop. A similar radial trend has been reported for the hot gas metallicity in some galaxy clusters.

{\tt Conclusions.} 
The ionization mechanisms of cooling flow elliptical galaxies are diverse, suggesting multiple channels for powering the warm ionized gas. {The positive correlation found in warm and hot gas metallicities suggest the intimate connection between the two gas phases, likely driven by gas cooling and/or mixing.} The large variation in the warm gas metallicity further suggests that cold gas mass derived under the assumption of solar metallicity for the CO-to-H$_2$ conversion factor needs to be revised by approximately an order of magnitude. 
}
   \keywords{Elliptical galaxies -- Metallicity -- AGN feedback}
 
  \titlerunning{Warm and Hot Gas Metallicity in Massive Elliptical Galaxies and AGN Feedback}
  \authorrunning{Olivares et al.}
    \maketitle

%%%%%%%%%%%%%%%%%%%%%%%%%%%%%%%%%%%%%%%%%%%%%%%%%%
%%%%%%%%%%%%%%%%% BODY OF PAPER %%%%%%%%%%%%%%%%%%

\section{Introduction}

    Active Galactic Nucleus (AGN) feedback may play a crucial role in quenching star formation and enriching the surrounding medium with metals through expanding X-ray cavities via jets and outflows in the surrounding hot halo 
  \citep{mcnamara00,gitti10,fabian12,olivares22b,gaspari20}. The clearest observations of such AGN feedback mechanisms appear in massive elliptical galaxies at the centers of clusters and groups. These galaxies typically show extended multiphase filaments—a mix of inflowing and outflowing gas forming a circulating flow. These filaments likely emerge from thermally unstable cooling of the hot intracluster medium (ICM), triggered by AGN feedback \citep{mcdonald10,mcdonald12,conselice01,hamer16,salome03,salome11,edge01,russell19,vantyghem19,temi18,Jimenez-Gallardo2021,Vigneron2024,Ubertosi2023,olivares23,Prathamesh2023,Gingras2024,Olivares2025}. Metal content in multiphase gas, as an important component of this ``baryon cycle'', remains poorly understood.
    
    The abundance of the X-ray emitting hot gas has been studied for different chemical elements at the center of clusters and some massive groups \citep{Su2013}. X-ray observations from \textit{Chandra} and XMM-Newton satellites show that the intracluster medium and intragroup medium 
    is rich in metals that are synthesized by supernovae (SNe) explosion    \citep{mernier16,mernier17,mernier18,dePlaa2017,Fukushima23}. Studies of central cluster galaxies show that X-ray cavities formed by AGN feedback can lift metals in their wake \citep{Kirkpatrick_2009}. This process creates higher hot gas-phase metallicity along the jet axis compared to the perpendicular direction \citep{liu19}—a finding also supported by hydrodynamical simulations of AGN outflows/jets (\citealt{gaspari11a,gaspari11b}). Yet questions remain about how different gas phases interact and how AGN feedback influences metal distribution around the central galaxy.

    Despite the rich availability of new deep integral field spectroscopy (IFU) observations \citep[e.g.,][]{olivares19,tremblay18}, measurements of the metallicity of the colder phase of the gas, e.g., warm-phase gas, are extremely limited, with only a handful of objects derived through Oxygen abundance (O/H) measurements \citep{Lagos2022,Ciocan21}. 
    
    Measuring metallicity is crucial for calculating accurate cold gas masses from CO emission lines, since the CO-to-H2 conversion factor depends heavily on it \citep{Bolatto13}. To date, all ALMA studies of massive cluster galaxies \citep{salome11,olivares19,russell19} have assumed solar metallicity. This assumption may underestimate the cold gas mass by up to an order of magnitude if the actual cold gas metallicity is 0.5~Z$\odot$ \citep{Amorin2016}. Yet measurements of metallicity in the colder phases of these massive systems remain virtually non-existent.
    
    In this paper, we present metallicity measurements of warm and hot gas phases in 13 massive  galaxies located in cluster and group centers using MUSE/VLT and \textit{Chandra} observations. These measurements help us understand the connection between different gas phases and the effect of AGN feedback on metallicity. In Section~\ref{sec:sample}, we describe our sample sources. In Section~\ref{sec:data}, we detail our observations. Section~\ref{sec:analysis} explains our metallicity measurement methods for both temperature gas phases. We present our findings in Section~\ref{sec:results}, discuss them in Section~\ref{sec:discussion}, and summarize our conclusions in Section~\ref{sec:summary}.

\section{Sample}
\label{sec:sample}

        \begin{table*}
        \caption{Summary of the sample}
            \centering
            \begin{tabular}{lcccc}
            \hline\hline
            {Source name} & {$z$} & SFR & $\rm M_{\star}$ & References\\
             & & ($\rm M_{\odot}\, yr^{-1}$) & ($\rm 10^{11}\, M_{\odot}$) & \\
             \hline
            2A0335+096 & 0.03634 & 0.46$\pm$0.66 &  5.33$\pm$0.26 & (1,2)  \\
            Abell\,2597 & 0.08210 & 5 - 12 & 3.24$\pm$0.31 &  (1,2)\\
            Abell\,S1101 & 0.05639 & 1.02$\pm$2.57& 6.13$\pm$0.55&  (1,2) \\\
            Abell\,1795 & 0.06326 &  3.45$\pm$5.01 &  6.99$\pm$0.52&  (1,2) \\  
            Abell\,1835 &  0.25200 &  117.48$\pm$1.58, 138 & 5.7 - 7.69  & (1,2)\\
            Abell\,496 & 0.03301 & .. & 1.85 & .. \\ 
            Centaurus (NGC4696)& 0.01016 &  0.2$\pm$0.1 & 6.15$\pm$0.14  & (1,2) \\
            Hydra$-$A & 0.05435 & 8$\pm$7 - 18  & 4.18$\pm$0.24 &  (1,2) \\
            M87 (NGC4486)& 0.00436 & 0.14$\pm$0.07 & 3.98& ..\\
            M84 (NGC4374) & 0.00339 & .. & 3.85& .. \\
            NGC5846 & 0.00571 & 0.019$\pm$0.003 & 0.67 & (3) \\ 
            PKS0745$-$19 & 0.10280 &  13.4$\pm$1.73,  24  & 5 & (1) \\ 
             RXJ0821+0752 & 0.10900 & 36.30 &  1.4 & (1) \\
            \hline
        \end{tabular}
        \tablebib{
        (1) \citet{mittal15}; (2) \citet{tremblay15}; (3) \citet{osullivan18a}}
        \label{tab:sample}
        \end{table*}
        
    As a pilot project we selected massive elliptical galaxies for this study based on existing deep MUSE and \textit{Chandra} observations. The majority of our sources came from \citet{olivares19} and consist of well-studied central cluster galaxies, including the notable Abell\,1835. We also included NGC5846, a central group galaxy from \citet{olivares22a,Lagos2022,Loubser2022}, and M84, an isolated elliptical galaxy from \citet{Eskenasy2024,Eskenasy2025}. Abell\,496 was an additional source studied in \citet{olivares23} and \citet{Ubertosi2024}. Our sample encompasses diverse sources with varying properties: star formation rates (between 0.001 and $>$100\,M${\odot}$\,yr$^{-1}$), stellar masses (0.6 -- 7$\times$10$^{11}$M$\odot$), H$\alpha$ morphologies, and redshifts from $0.00339$ to $0.2500$. Table~\ref{tab:sample} lists some of these properties.
    
\begin{figure*}
        \centering
        \includegraphics[width=2.1\columnwidth]{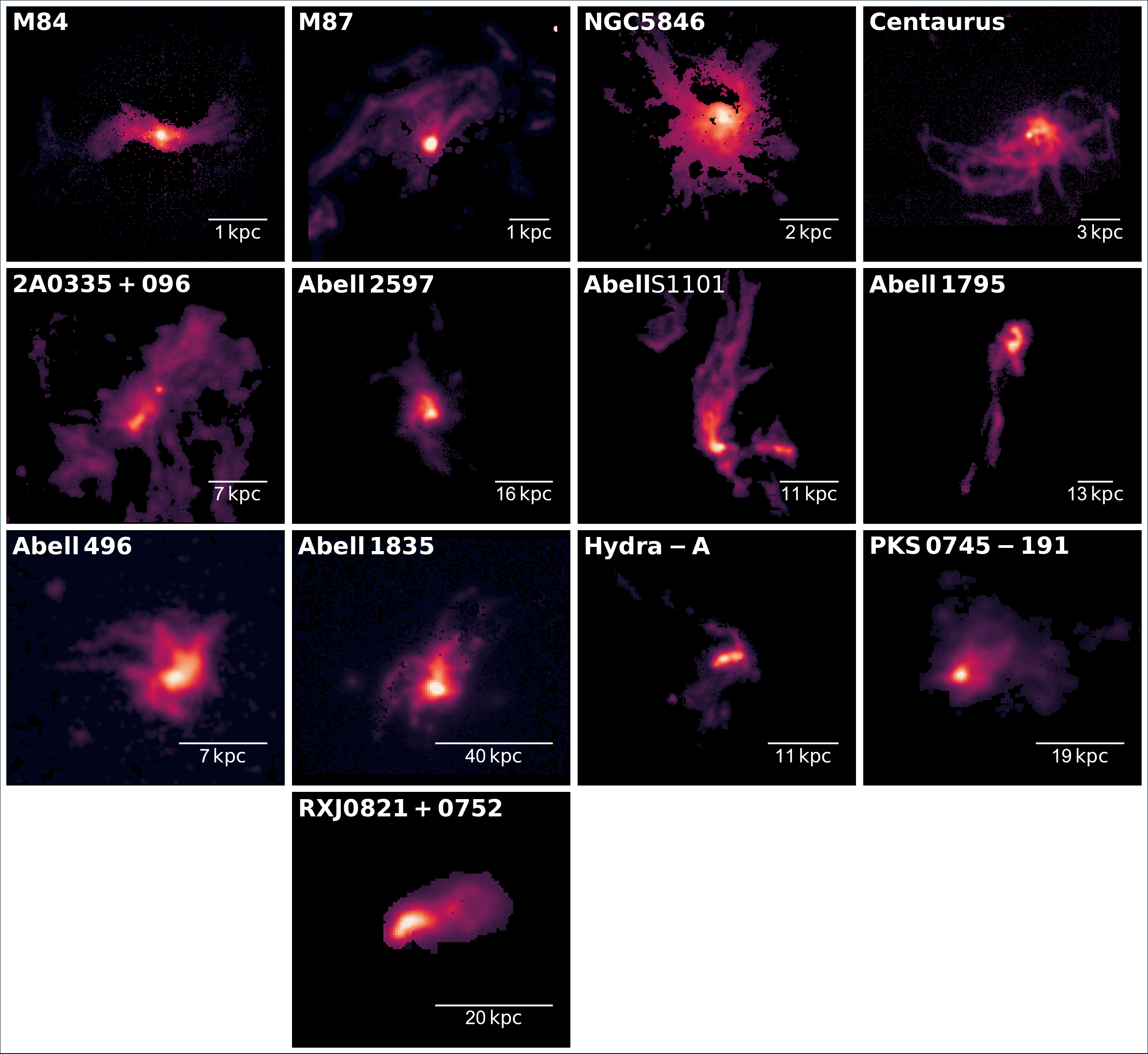}\\
        \caption{Map of warm ionized gas traced by H$\alpha$ emission from MUSE observations of our sample.} \label{fig:sample}
        \end{figure*}

    Figure~\ref{fig:sample} shows images of the H$\alpha$ emitting gas for each source in our sample, demonstrating their diverse structures and sizes. All sources have clear indication of AGN radio feedback through either the detection of radio jets or X-ray cavities. All the sources have filamentary structures at different scales.
    
\section{Observations and data reduction}
\label{sec:data}

\subsection{MUSE observations}

    We analyzed archival MUSE observations from our sample sources to measure the optical properties of the warm ionized gas. Table~\ref{tab:observations} presents the observational properties of the MUSE data for each cluster.
    
    MUSE is an optical integral field spectrograph mounted on the Very Large Telescope (VLT) in Chile. It has a $1\arcmin\times1\arcmin$ field of view with a pixel size of $0.2\arcsec$, and detects wavelengths from 4750$\AA$ to 9350$\AA$. All observations were conducted in wide field mode (WFM) with a spectral resolution of $R=3000$. Exposure times varied by source, ranging from 2,700 to 11,010 seconds.
    
    We processed the MUSE data using MUSE pipeline version 1.6.4 and ESOREX 2.1.5, which performed standard reduction of individual exposures and combined them into a final datacube. Beyond the sky subtraction provided by ESOREX, we applied additional sky subtraction using ZAP (Zurich Atmosphere Purge; \cite{soto16}).
    
    We modeled the final data cube using Penalized Pixel Fitting {\sc pPXf} method \citep{Cappellari2023}. We modeled the stellar continuum using E-MILES libraries \citep{Vazdekis2016}. We fitted the stellar and emission lines together. Each optical emission line—such as [OII]\,$\lambda\lambda$3726,3729, [OIII]\,$\lambda$4959,[OIII]\,$\lambda$5007, [OI]\,$\lambda$6300, [OI]\,$\lambda$6363, [NII]\,$\lambda$6584, [NII]\,$\lambda$6548, [SII]\,$\lambda$6716, [SII]\,$\lambda$6731, [NI]\,$\lambda$5002, [CaII]\,$\lambda$7291, [CaII]\,$\lambda$7323, H$\alpha$, H$\beta$, HeII\,$\lambda$4686, among others, were fitted with a Gaussian component. We fitted the doublets, such as [NII] and [OII] with separated gaussian and without assuming a theoretical ratio. Two Gaussian were used for some sources in the central region where the AGN is located and a broader component was found. The velocities and velocity dispersions of all emission lines were tied together to ensure a consistent kinematic solution. Figure~\ref{fig:spectrum} in Appendix~\ref{appendix:spectrum}  presents an example of the best-fit of the best-fit model overlaid on the observed optical MUSE spectrum. 
    
    To correct for Galactic foreground extinction, we used the recalibration of the \cite{schlafly11} dust map of the Milky Way, which is based on IRAS+COBE data \cite{schlegel98}. We generated spatially resolved flux maps of relevant emission lines for our study, including H$\alpha$, H$\beta$, [OI], [SII], [NII], [OIII] doublets and auroral lines.
    
    We corrected for intrinsic extinction in each spaxel using the colour excess $\rm E(B-V)$ from the gas, based on the Balmer decrement H$\alpha$/H$\beta = 2.86$, which corresponds to case B recombination (\citealt{OsterbrockFerland2016}, with $\rm n_{e} =100$ cm$^{-3}$ and $\rm T_{e}=10,000$ K). We applied the \citet{cardelli89} extinction curve with $\rm R_{V}=3.1$.
    
\subsection{\textit{Chandra} X-ray observations}

    We used all available archival \textit{Chandra} observations for each cluster, except for Abell 1795, where we selected a subset of data to maintain consistent count numbers across sources. We performed data reduction and calibration using \textit{Chandra} Interactive Analysis Observations software (CIAO) 4.16 \cite{fruscione06} and \textit{Chandra} Calibration Database (CALDB) 4.9.2.1. Using the chandra\_repro tool in CIAO, we conducted standard calibration and data reprocessing. We then subtracted standard blank sky background files and readout artifacts. Using CIAO's wavdetect package, we identified and removed point sources from the exposure-corrected \textit{Chandra} image. Table~\ref{tab:observations} summarizes the observational properties of the \textit{Chandra} observations for each cluster.

\section{Analysis}\label{sec:analysis}

    Warm-gas phase abundance measurements are highly dependent on the process that ionized the gas. The broad wavelength coverage of the MUSE spectra enables comparison of several optical emission-line diagnostics to constrain the ionization sources of the warm gas, which is necessary to determine the Oxygen abundance of the warm ionized gas. The primary optical line ratios used to study the gas ionization mechanism are [NII$]\lambda$6583/H$\alpha$ and [OIII]$\lambda$5007/H$\beta$, known as the Baldwin, Phillips, \& Terlevich diagram (BPT; \citealt{Baldwin81}). We also analyze the warm ionized gas line ratios [SII]/H$\alpha$ and [OI]/H$\alpha$, as these ratios are sensitive to the source of ionizing radiation.
    
    The BPT-diagrams are useful to discriminate between the low-ionization process of the gas due to stellar photoionization (also known as star-forming HII regions) and due to harder ionizing processes such as those produced from the central AGN and fast shocks. We used the spatially resolved BPT diagram to separate the different regions, namely, HII region, AGN, LINERS (Low-ionization nuclear emission-line region)\footnote{Note that the LINER term is not appropriate in this context since LINERs was originally defined as being ``nuclear" emission (\citealt{heckman80,ho93}; alternately, LIER (low ionization emission-line regions) has been suggested by recent authors.} (low ionization nuclear emission) and composite region. We caution that the classifications can be diagram-dependent, for example a region can be classified as a LINER-like in the SII-BPT or OI-BPT but as a star formation on the NII-BPT diagram. 
        
\section{Results}\label{sec:results}
\subsection{Ionization mechanism of the warm gas}

    Before computing the O/H abundance of the warm gas, one needs to understand what is ionizing the gas in order to use a proper abundance calibrator. The BPT-diagram can be used to constrain the main source of ionizing radiation, by classifying a set of emission lines into one of the four classes: (i) HII regions, (ii) Seyfert, where the ratios are consistent with ionization by AGN, (iii) LI(N)ER, where the ratios are consistent with multiple mechanism -- photoionization by hot and evolved starts, pAGB stars, shock-excited gas, reprocessing of the EUV and soft X-ray radiation from the plasma cooling, and mixing layers, (iv) composite of several ionization processes, {as the ones mentioned before}.

    For our sample, we identify three types of systems based on the BPT diagram - i) The ones that only occupies LI(N)ER-like emission region (Centaurus, M87, M84, Abell\,496), ii) sources with most of the spaxels in the composite region (Abell\,2597, PKS0745$-$19, Abell S1101, Abell\,1835, Abell\,1795, 2A0335+096, Hydra-A, NGC5846), and iii) sources with most of spaxels in the HII region, but also in the composite (RXJ0821+0752). 
    
    The central regions of most sources are dominated by LINER/LIER ionization (PKS0745$-$19, Abell\,S1101, 2A0335+096 in NII-BPT). RXJ0821+0752 presents an interesting case: its central region shows HII-dominated ionization, while regions at greater distances exhibit composite ionization.

    We find that  the ionization sources in our sample  have variable impacts at different radii \citep{Lagos2022}. Therefore, the determination of Oxygen abundances cannot rely on the assumption of classical HII photoionization conditions throughout the interstellar medium of the galaxies. 
    
        \begin{figure}
        \centering
         \includegraphics[width=\columnwidth]{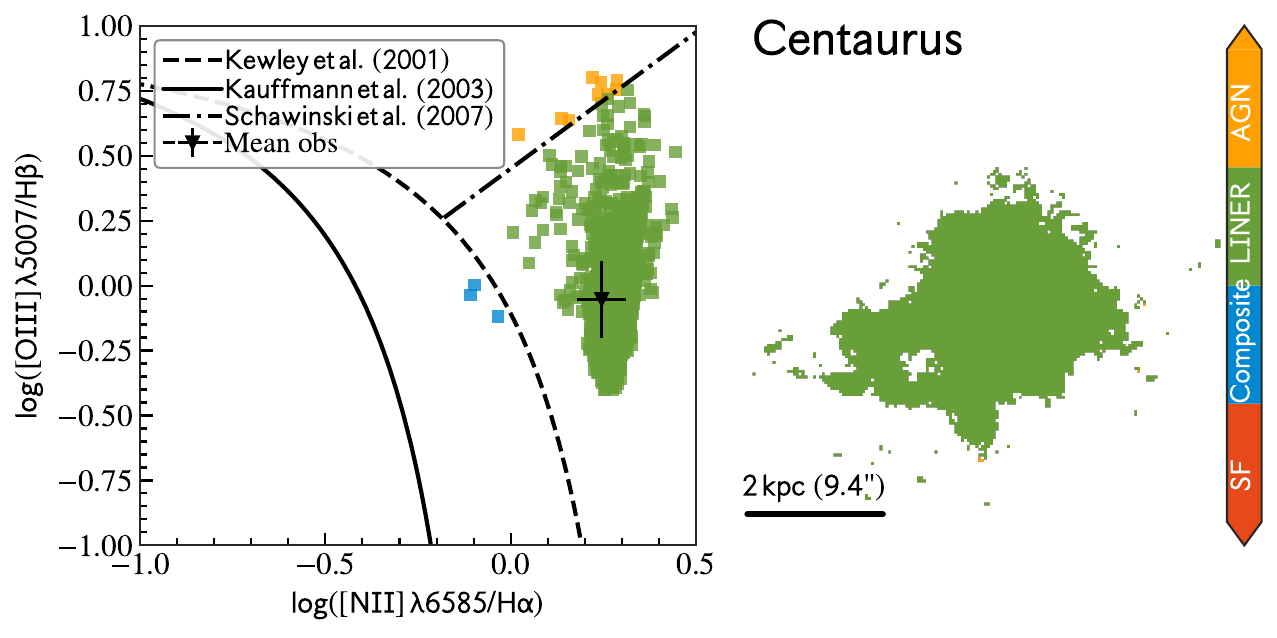}\\
         \includegraphics[width=\columnwidth]{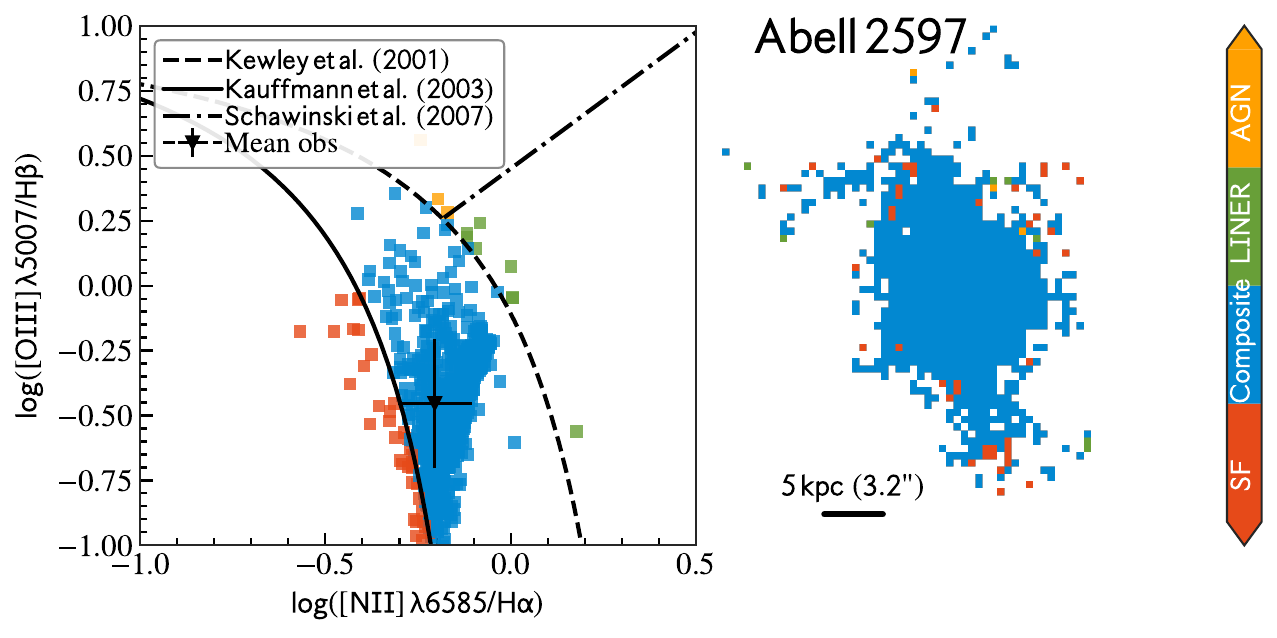}\\
         \caption{Example of resolved BPT-NII classic diagrams for Centaurus (top panels), and Abell\,2597 (bottom panels). Spaxels are color-coded based on their location relative to boundaries between the well-known empirical and theoretical classification schemes \citet{kewley01,kauffmann03,schawinski07} shown in black solid, dashed, and pointed lines, respectively. Yellow, green, blue, and red spaxels correspond to AGN-, LI(N)ER-, composite-, and HII-dominated regions, respectively.}
         \label{fig:BPT}
        \end{figure}
        
\subsection{Abundance of the warm gas}
\label{sec:metallicity_warm}

\setlength{\tabcolsep}{4pt}
\renewcommand{\arraystretch}{1.1}
        \begin{table*}
            \caption{Summary of metallicity measurements}
            \centering
        \begin{tabular}{lcc|cccccc}
               \hline
               \hline
         & \multicolumn{2}{c}{Z$_{\rm Hot}$\,/\,Z$_{\odot}$} & \multicolumn{6}{c}{Z$_{\rm Warm}$\,/\,Z$_{\odot}$} \\
        Method &   &  &  AGN & HII  & LI(N)ER & Te & X-ray & pAGB  \\
        Source &  &   &  &  &  & &  &  \\
        (a) &  &  (b) &  (c)& (d) &(e) &(f)  &(g)  & (h) \\
        \hline
        2A0335\_1 & .. & 1.05$\pm$0.18 & .. & .. & 1.4$\pm$0.03 & .. & 0.82$\pm$0.04 & 1.1$\pm$0.15 \\
        2A0335\_2 & .. & 1.05$\pm$0.18 & .. & .. & .. & .. & 0.76$\pm$0.01 & 0.98$\pm$0.21 \\
        A1795\_1 & .. & 0.92$\pm$0.1 & .. & .. & .. & .. & 0.58$\pm$0.03 & 0.95$\pm$0.21 \\
        A1795\_2 & .. & 0.9$\pm$0.08 & .. & .. & .. & .. & 0.59$\pm$0.11 & 0.85$\pm$0.2 \\
        A1835 & .. & .. & .. & .. & .. & 0.50$\pm$0.06 & .. & .. \\
        A1835\_1 & .. & 0.67$\pm$0.17 & .. & .. & .. & .. & 0.47$\pm$0.01 & 0.45$\pm$0.15 \\
        A1835\_2 & .. & 0.58$\pm$0.08 & .. & 0.87$\pm$0.06 & .. & .. & 0.72$\pm$0.08 & 0.31$\pm$0.15 \\
        A2597\_1 & .. & 0.61$\pm$0.04 & .. & .. & 1.29$\pm$0.02 & .. & 0.51$\pm$0.01 & 0.56$\pm$0.06 \\
        A2597\_2 & .. & 0.42$\pm$0.02 & .. & .. & .. & .. & 0.42$\pm$0.01 & 0.68$\pm$0.18 \\
        A496\_1 & .. & 1.29$\pm$0.08 & .. & .. & 1.43$\pm$0.02 & .. & .. & 1.26$\pm$0.2 \\
        A496\_2 & .. & 1.29$\pm$0.08 & .. & .. & 1.44$\pm$0.04 & .. & .. & 1.51$\pm$0.17 \\
        AS1101\_1 & .. & 0.7$\pm$0.07 & .. & .. & 1.31$\pm$0.02 & .. & 0.77$\pm$0.01 & 0.98$\pm$0.17 \\
        AS1101\_2 & .. & 0.64$\pm$0.11 & .. & .. & .. & .. & 0.59$\pm$0.01 & 0.85$\pm$0.2 \\
        Hydra-A\_2 & .. & 0.76$\pm$0.08 & .. & .. & 1.02$\pm$0.03 & .. & 0.55$\pm$0.07 & 0.98$\pm$0.21 \\
        M84\_1 & .. & 1.26$\pm$0.02 & .. & .. & 1.31$\pm$0.06 & .. & .. & 1.51$\pm$0.29 \\
        M84\_2 & .. & 1.26$\pm$0.02 & .. & .. & 1.36$\pm$0.05 & .. & .. & 1.12$\pm$0.2 \\
        M87\_1 & .. & 1.25$\pm$0.02 & .. & .. & 1.41$\pm$0.04 & .. & .. & 1.51$\pm$0.24 \\
        M87\_2 & .. & 1.25$\pm$0.02 & .. & .. & 1.28$\pm$0.04 & .. & .. & 1.51$\pm$0.24 \\
        NGC5846\_1 & .. & 0.39$\pm$0.15 & .. & .. & .. & .. & 0.58$\pm$0.29 & 0.63$\pm$0.18 \\
        NGC5846\_2 & .. & 0.39$\pm$0.15 & .. & .. & 1.29$\pm$0.02 & .. & 0.46$\pm$0.35 & 0.43$\pm$0.13 \\
        PKS0745\_1 & .. & 0.66$\pm$0.07 & .. & .. & .. & .. & 0.72$\pm$0.02 & 0.98$\pm$0.19 \\
        PKS0745\_2 & .. & 0.72$\pm$0.1 & .. & .. & .. & .. & 0.75$\pm$0.08 & 0.95$\pm$0.21 \\
        RXJ0821 & .. & .. & .. & .. & .. & 0.30$\pm$0.06 & .. & .. \\
        RXJ0821\_1 & .. & 0.59$\pm$0.08 & .. & 0.89$\pm$0.07 & .. & .. & 0.49$\pm$0.06 & 0.68$\pm$0.2 \\
        RXJ0821\_2 & .. & 0.59$\pm$0.08 & .. & .. & .. & .. & 0.72$\pm$0.07 & 0.78$\pm$0.2 \\
        Centaurus\_1 & .. & 1.34$\pm$0.06 & .. & .. & 1.4$\pm$0.02 & .. & .. & 1.51$\pm$0.08 \\
        Centaurus\_2 & .. & 1.31$\pm$0.07 & .. & .. & 1.41$\pm$0.02 & .. & .. & 1.51$\pm$0.11 \\
        \hline
        \hline
        \end{tabular}
            \tablefoot{
            \label{tab:metallicity}
        {(a) Source names together with the region ID. If the source doesn't have an ID region, it means the metallicity was derived using the integrated flux.}\\
        {(b) Metallicity of the hot-phase gas computed as described in Sec.~\ref{sec:hot_abundance} for different regions.}\\
        {(c) Warm-phase gas metallicity derived using eq.~\ref{eq:Marino2013} within the HII-emitting region.}\\
        {(d) Warm-phase gas metallicity derived using eq.~\ref{eq:kumari} within the LI(N)ER-emitting region.}\\
        {(e) Warm-phase gas metallicity derived using the $T_{\rm e}$-based method as described in Sec.~\ref{sec:Te}.}\\
        {(f) Warm-phase gas metallicity derived X-ray/EUV models described in Sec.~\ref{sec:polles21}.}\\
        {(g) Warm-phase gas metallicity derived pAGB stars models described in Sec.~\ref{sec:polles21}.}\\
        }
        \end{table*}
        
    The gas-phase Oxygen abundance (O/H) is a widely used and reliable tracer of metallicity in the interstellar medium of galaxies. Oxygen is one of the most abundant heavy elements, and its key ionization stages produce strong optical emission lines, making it readily accessible through spectroscopic observations. We derived the gas-phase O/H abundances using emission-line ratios from {MUSE data}, applying a range of calibration methods tailored to the dominant ionization mechanism in each source and region. These include: (1) the HII-region calibration from \citet{Marino2013}, (2) the LI(N)ER calibration from \citet{Kumari2019}, and models appropriate for composite and LI(N)ER-like ionization sources, such as (3) X-ray/EUV photoionization models from \citet{polles21}, and (4) post-Asymptotic Giant Branch (pAGB) star models from \citet{rauch2003}. Additionally, we computed metallicities using the direct $T_{\rm e}$-method where applicable. A detailed comparison of these methods and their associated systematics is presented in Appendix~\ref{appendix:warm_phase}.

\subsubsection{HII ionization}\label{sec:HII}
    For sources showing HII ionization, we derive the O/H gas abundance the ${O⁢3⁢N⁢2}=\mathrm{log⁡([O\,III]⁢\lambda5007/H⁢\beta)\, - \,log([\,NII]⁢\lambda6583/H⁢\alpha)}$ calibrated by \citet{Pettini2004}, and later refined by \citet{Marino2013}. 
    \begin{equation}
        \mathrm{12 + log(O/H)} =\mathrm{8.533\pm0.012 -(0.214\pm0.012) \cdot O3N2}\\
    \end{equation}\label{eq:Marino2013}
    This calibration is valid for ${O3N2}$ values between $-$1.1 and 1.7, which approximately correspond to an Oxygen abundance range of $\mathrm{12 + \log(O/H)} = 8.16$ to $8.76$.
        
\subsubsection{Composite and LI(N)ER ionization}
\label{sec:polles21}
    Two mechanisms that have been described as potential responsible for powering the ionization of the filaments, composite and LI(N)ER-like ionization, are the soft X-ray and EUV emitting from the cooling gas \citep{ferland09,polles21}, and pAGB stars \citep{Binette94,Sarzi10,Lagos2022}. In particular soft X-ray/EUV are the great interest, since the multiphase filaments observed at the center of BCGs are believed to be form through the condensation of the hot ICM \citep[e.g.,][]{li20,voit19,gaspari20}.\\
    
    \paragraph{X-ray and EUV models:} \citet{polles21} present predicted emission-line ratios from CLOUDY models, described as single-component plane-parallel models of constant-pressure self-irradiated clouds. In \citet{polles21} models, the emission lines are photoionized by X-ray/EUV emission and can reproduce well the optical emission lines observed in cooling clusters. The models also included extra heating described with a turbulent velocity, which can be produced by, e.g., AGN jets, turbulent mixing between hot and cold phases, and collisions between filaments. The models incorporate a range of metallicities ($Z/Z_{\odot}$) (between 0.3 and 1$Z_{\odot}$), and turbulent velocities (Between 0.0 and 100\,km\,s$^{-1}$).
    We derived the O/H abundance of the warm-phase gas using grids of models present in \citet{polles21}. 
    
    Some sources in the LI(N)ER region with very high [NII]/H$\alpha$ ratios, including Centaurus, M87, and A\,496, could not be fitted with \citet{polles21} models because of their high metallicity ($Z>1Z_{\odot}$). \\

    \paragraph{pAGB models:} We derived the O/H abundance using the pAGB NLTE models \citep{rauch2003} implemented on the HII-CHI-mistry code for optical emission lines v.5.5 \citep{Perez-Montero2019}. HII-CHI-Mistry, which was originally developed by \citet{Perez-Montero2014} for HII regions but was later extended to AGN sources \citep{Perez-Montero2019} and extreme emission-line galaxies \citep{Perez-Montero2021}. HII-CHI-Mistry is Bayesian-like Python, which was originally developed by to calculate chemical abundances and physical properties using emission line fluxes from ionized gaseous nebulae. HII-CHI-Mistry employs a grid of photoionization models with three free parameters: the chemical properties of the gas-phase ISM, 12+log(O/H) and log(N/O), and the ionization parameter log(U). These parameters are estimated from emission line ratios that are sensitive to them.    
    
    As input, we used the following emission lines, depending on the source, [OIII]$\lambda 4959$, [OIII]$\lambda 5007$, [NII]$\lambda 6584$, [SII]$\lambda 6717$, and [SII]$\lambda 6731$, and [OII]$\lambda$7319,7330. 
    
\subsubsection{LI(N)ER ionization}\label{sec:kumari}

    For LI(N)ER dominated regions,
    we used the calibration presented in \citet{Kumari2019}, as described below.
    \begin{equation}\label{eq:kumari}
    \small \mathrm{12 + log(O/H)} = \mathrm{7.673 + 0.22 \cdot \sqrt{25.25 - 9.073 \cdot O3N2} + 0.127O3}
    \end{equation}
    where, ${O3}=\mathrm{log([O\,III]\lambda5007/H\beta)}$.

\subsubsection{Direct $T_{\rm e}$-method}\label{sec:Te}

    For Abell\,1835 and RXJ0821+0752, we calculated the O/H abundance using the $T_{\rm e}$ method \citep{Kewley08}. We used the integrated [OII]$\lambda3726+\lambda3729$ luminosities from \citep{Gingras2024} detected with Keck Cosmic Web Imager (KCWI). We recall that adopting a constant F([OII]$\lambda$3727)/F(H$\beta$) ratio results in oxygen abundance estimates with negligible variation: using F([OII]$\lambda$3727)/F(H$\beta$) between 0.3–0.8 yields a (O/H) difference of 0.03 dex \citep{LagosP2014}. To ensure consistency, we used the same aperture to measured [OIII]$\lambda$4959, [OIII]$\lambda$5007 and H$\beta$ fluxes from the MUSE observations for both sources. 
    
    We derived an integrated $T_{\rm e}$ and density, $n_{\rm e}$, using pyNeb \citep{Luridiana2015}, along with the [NII]$\lambda$6548, [NII]$\lambda$6584, and [NII]$\lambda$5755 auroral lines across the nebulae, and the [SII] emission lines to determine the electron density, $n_{\rm e}$. We also tried [OIII]$\lambda$4636 and [OIII]$\lambda$5007 ratios to derive the electron $T_{\rm e}$ for Abell\,1835. We found that [OIII] lead to higher $T_{\rm e}$ compared to [NII] ratios. [OIII]$\lambda$4636 was not detected for RXJ0821+0752. This behavior has been reported in the literature \citep[e.g.,][]{Khoram2024}, where several line ratios were used to derive $T_{\rm e}$, and it appears that [OIII] is the only indicator that leads to a higher $T_{\rm e}$ at higher metallicity ($>8.6$) due to likely contamination from Fe$\lambda$4360 or an unknown physical process (Fig.5 of \citealt{Khoram2024}). {We carefully check the [OIII] auroral line, indicating contamination from Fe$\lambda$4360.} Due to that reason, we used the $T_{\rm e}$ derived using [NII] ratio. For Abell\,1835, we found a $T_{e}$ of 8888 $\pm$ 164. For RXJ0821+0752, we derived a $T_{e}$ of 8934 $\pm$ 251.
    
    Following, we derived the Oxygen abundance using the {\sc atom.getIonAbundance} routine from pyNeb, and the line radios, R2 = ([OII]$\lambda$3726+[OII]$\lambda$3729)/H$\beta$, and R3= ([OIII]$\lambda$4959+[OIII]$\lambda$5007)/H$\beta$ to derive the $\frac{O^{+}}{H^{+}}$ and $\frac{O^{++}}{H^{++}}$, respectively.
    
    Then the total O/H abundances were derived using,
    \begin{equation}
      \rm \frac{O}{H} = \frac{O^{+}}{H^{+}} + \frac{O^{++}}{H^{++}}
    \end{equation}
    
    In Appendix~\ref{appendix:warm_phase}, we present a comparison between the $T_{\rm e}$ abundance measurements and alternative methods.

\subsection{Abundance of the hot gas}\label{sec:hot_abundance}

    We extracted spectra from roughly the same region as that covering the warm gas to determine the abundance of the hot gas. For some sources we used multiple regions, depending on the amount of counts.
    The spectra were fitted using XSPEC version 12.13.1 \citep{arnaud1996xspec} to derive abundance using  two {\tt apec} models, which represent thermal emission of the hot ICM gas and the cooler emitted gas ({\tt phabs*}({\tt apec}+{apec})). The fit was restricted to 0.5-7.0~keV energy band. A Blank-Sky field was used to constrain the background emission. The adopted values of the Galactic column density ($N_{\rm H}$) for the galactic absorption ({\tt phas}) model were obtained from the Colden (Galactic Neutral Hydrogen Density Calculator). We used the solar abundance table from \citet{Asplund2009}. {In the Appendix Fig.~\ref{fig:spectrum} we show an example of X-ray spectrum of one region in Abell 2597 using one observation, together with the best fit model.}
    
    The abundance of the two {\tt apec} models were tied together while the temperature and the normalization were allow to vary freely. In Table~\ref{tab:metallicity} we list our results. For the case of NGC5846, we fixed the lower temperature to half of the higher temperature ($\rm kT_{high} = 0.93\pm 0.02$ keV). Our results are consistent with previous works \citep[e.g.,][]{Panagoulia2015}.
    
\section{Discussion}\label{sec:discussion}

    This paper investigates the metallicity of both the warm and hot gas phases in massive elliptical galaxies. Because the ionization mechanism of the warm gas is uncertain, we derived gas-phase abundances using multiple approaches. For LI(N)ER-dominated regions, we applied Eq.\ref{eq:kumari} and post-AGB (pAGB) models. For HII regions, we used Eq.\ref{eq:Marino2013} to estimate the O/H abundance. For composite regions, we derived the O/H abundances of the warm gas using pAGB and X-ray/EUV photoionization models, which account for the composite nature of the ionization in these regions.
    
    Subsequently, we discuss our results in the context of the AGN feedback mechanism. We then compare the warm and hot phase metallicity of our sample and examine the implications of our findings. We assume a solar O/H abundance of 8.69 from \citet{Asplund2009}.
    
\subsection{AGN feedback effect on the O/H abundance}
\label{sec:abundace_maps}
        
    Figures~\ref{fig:abundance_maps} and \ref{fig:abundance_maps2} show median O/H abundance profiles of the warm gas. The resolved BPT classification, color-coded by ionization mechanism, is also shown.  Additionally, we provide resolved O/H abundance maps, derived using the calibration or model appropriate for each spaxel’s ionization source.  Maps for the remaining sources are presented in Appendix~\ref{fig:app_abundance_maps1}, \ref{fig:app_abundance_maps2}, \ref{fig:app_abundance_maps3}, \ref{fig:app_abundance_maps4}, and \ref{fig:app_abundance_maps5}.
    
    The HII calibrations presented in Section \ref{sec:HII} were applied exclusively to spaxels in the outer regions of Abell 1835 and the central region of RXJ0821, where the ionization is dominated by star-forming activity.
    
    Given the uncertainty in the dominant ionization mechanism within the composite spaxel, we derived the O/H abundance using two independent photoionization models: post-AGB and X-ray/EUV. Both models have been shown to reproduce the optical emission-line ratios observed in systems with mixed ionization (\citealt{Krabbe2021, polles21, Lagos2022}). Consequently, O/H abundances estimated from post-AGB and X-ray/EUV models are considered more robust, particularly in environments where the ionization source is ambiguous or comprises a mixture of contributors.
     
    For LI(N)ER spaxels, we also measured the O/H abundance using two independent methods: the calibrations presented by \citet{Kumari2019} (see Sec. \ref{sec:kumari}) and the post-AGB photoionization models, as they can also reproduce the LI(N)ER ionization (see \citet{Krabbe2021, Lagos2022}). 

    For each composite or LI(N)ER spaxel, we therefore computed two separate metallicity estimates, one from each model, without selecting between them. The O/H maps derived for each sources are shown in Figures~\ref{fig:abundance_maps} and \ref{fig:abundance_maps2}.

    To construct the O/H profiles, we first computed the oxygen abundance for each spaxel using the calibration or model appropriate for its ionization mechanism (as described above). The annuli were centered on the galaxy nucleus, and the median O/H value was calculated from all spaxels at a given radius for each O/H map separately. The shaded pink band in the profiles represents the range of O/H values obtained from the different methods, as shown and described in Appendix Figure \ref{appendix:fig_comparison_profile}. This band illustrates that, while the absolute O/H values may vary slightly between methods, the overall radial trend remains unaffected by the adopted calibration.
    
    In the same vein, the radial profiles presented in this work include both determinations for the composite and LI(N)ER spaxels, with the differences between the methods reflected in the shaded regions. We followed this approach to account for the ambiguity in identifying the dominant ionization mechanism in these regions. Both models yield similar trends in the O/H abundance profiles, with the post-AGB models giving, on average, values approximately 0.15 dex higher than those derived from the X-ray models (see Appendix \ref{appendix:warm_phase} and \ref{appendix:fig_comparison_profile}), indicating the gradient remains robust against the adopted ionization model.  

    The role of the ionization parameter, $U$, in shaping the derived metallicities deserves careful consideration. In our analysis, variations in $U$ do not bias the metallicity estimates, particularly in the case of the pAGB models, which compare the observed emission-line ratios with grids of photoionization models to simultaneously determine ($\rm 12 + log10(O/H)$) and ($\rm log10 U$) in a non-degenerate way. In these models, $U$  is primarily constrained by the relative strength of the [NII] lines and high-excitation coronal lines such as [NII]$\lambda$5755, both of which are sensitive to the hardness of the ionizing radiation field. The resolved maps of both quantities reveal that regions with lower $\rm O/H$ also exhibit lower values of $\rm log10 U$, consistent with more efficient cooling and softer ionizing spectra in metal-rich environments. The observed gradient in $\rm log10 U$ supports the presence of a diluted and extended ionizing field—likely produced by pAGB or X-ray sources rather than a central AGN. Although we cannot entirely rule out that part of the O/H radial profile reflects model-dependent correlations (see \citet{ji2022}), variations in the ionization parameter are explicitly incorporated into the abundance determinations. Consequently, differences in excitation or radiation-field hardness are intrinsically accounted for in the derived metallicities.

    As shown in Figures~\ref{fig:abundance_maps} and \ref{fig:abundance_maps2}, some, but not all, sources in our sample exhibit clear positive O/H abundance gradients in the inner region of the gas, with O/H increasing with distance from the galaxy center. This behavior appears independent of the ionization of the source, and thus, the abundance calibrator or models. The O/H abundance difference between central and outer gas regions spans 0.1 to 0.4\,Z$_{\odot}$.
    
    X-ray observations of some galaxy clusters and groups show that overall the abundance is higher at the center, and then drops to larger radii (0.1--0.3\,$Z_{\odot}$) \citep[e.g.,][]{mernier17}. However, some galaxy clusters and groups show an abundance drop within 1–10 kpc in their hot gas profiles, with a peak at about 10 kpc before declining again at larger radii (e.g., Centaurus, M87, Abell\,2597, \citealt{liu19,cavagnolo09}). These central abundance drops in the hot gas appear across several elements — particularly Fe, Si, S, Mg, O, and Ca \citep[e.g.,][]{Churazov2003,Churazov2004,Panagoulia2015,mernier17,Gendron-Marsolais2017,Rasmussen2007}. These drops are more pronounced than those we observed in the warm gas phase, reaching from 0.5\,Z$_{\odot}$, to 2\,Z$_{\odot}$ in Centaurus within 10~kpc \citep{SandersFabian06}.
    
    One explanation for these central abundance drops is that a significant fraction of metals become deposited into dust grains and trapped in dusty filaments after cooling from the hot gas \citep[e.g.,][]{panagoulia13,Panagoulia2015,liu19}. These dust grains are then displaced by buoyant bubbles and jets, and released back into the hot gas outside the core, creating O/H abundance drops. This hypothesis predicts that abundance drops should not occur for two noble gases, Ne and Ar, as they do not easily deposit into dust grains \citep{Lakhchaura19}. 
    
    Another explanation for central abundance drops are due to mechanical processes, like the displacement of metal-rich gas from the center to larger distances by AGN-driven feedback \citep{liu19}. There are examples of central cluster galaxies where X-ray cavities and bubbles created by AGN feedback can lift metals in their wake. This leads to an enhanced hot gas-phase metallicity along the jet axis, compared to the perpendicular direction \citep{liu19}, as also supported by hydrodynamic simulations of AGN outflows \citep{gaspari11a}. In the same vain, MUSE observations of the Teacup (z=0.085) quasar also show higher O/H around the radio bubble edges, compared to rest of the nebulae, suggesting AGN feedback can produce metal enrichment at large distances \citep{Venturi2023}. Alternatively, the metals could be depleted in colder phases, such as the cold molecular gas.

    The observed O/H abundance drops could also be influenced by variations in the ionization parameter, as discussed above, as regions with higher excitation yield lower apparent abundance. While our modeling accounts for this degeneracy, we cannot fully rule out that part of the gradient arises from changes in ionization conditions. 
    
    We also note that the presence of a hidden AGN could not explain the large-scale (several kiloparsec) variations observed in the O/H abundance. Ionization from an AGN typically dominates only the innermost regions ($<$1 kpc).
    
    Overall, the variation of the abundance profile in the warm gas phase could be attributed to the contribution from these factors. This abundance drop may be closely linked to the abundance of the hot gas. It is worth mentioning that, as shown in X-ray studies, not all clusters show a warm-phase abundance drop. In fact, many clusters are consistent with flat warm O/H abundance profiles. One explanation is that as soon as the filaments form and cool down to lower temperatures ($<$10,000\,K), the hot gas is being replenished by the large scale flows.

\subsection{Warm and hot gas-phase metallicity correlation}

    \begin{figure}
    \centering
        \includegraphics[width=0.5\textwidth]{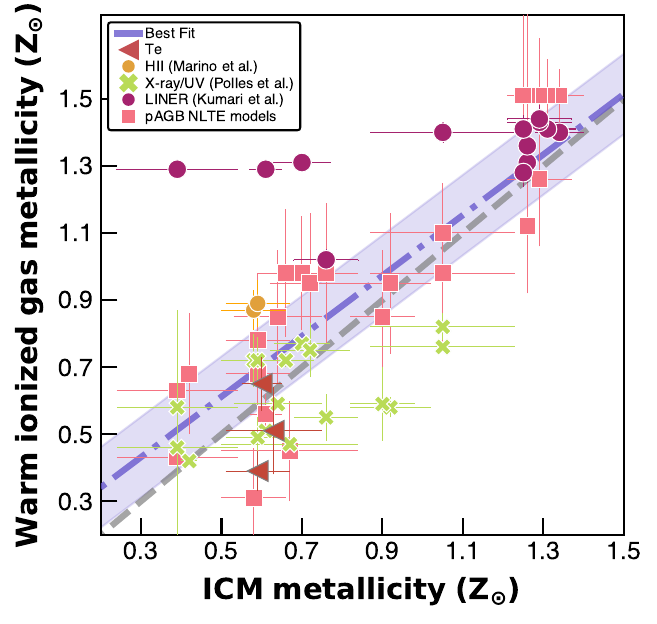}
        \caption{Comparison of warm and hot-gas phase metallicities for a sample of sources using MUSE and \textit{Chandra} observations. We derived warm gas metallicity using various calibration methods, depending on the source and their ionization mechanism. For LI(N)ER-like regions and sources (magenta circles), we used the \citet{Kumari2019} calibration. For HII-dominated regions, we used the \citet{Marino2013} calibration (yellow circles). Composite and LI(N)ER region derived using pAGB models are shown with pink rectangles. The warm-phase metallicity obtained from X-ray/EUV models are shown with green crosses. 
         Three sources, including MACS 1931.8-2635, have warm-gas metallicity measurements using the $\rm T_{e}$-direct method (red inverted triangles). 
         The errors for the warm gas phase abundance we included both the uncertainties associated with the calibration and 1$\sigma$ deviation for the regions. The purple dashed line correspond to the best fit, and the dashed region correspond to 1$\sigma$ uncertainty. The dashed gray line shows the one-to-one relation
         The best fit correspond to, $\rm Z_{warm} = (0.15\pm0.09) + Z_{hot}^{0.90 \pm 0.10}$.}
         \label{fig:warm_vs_hot_metallicity}
    \end{figure}

    We derived the metallicity of the warm gas by converting its O/H abundance using the solar O/H abundance of 8.69 from \citet{Asplund2009}, following,
    \begin{equation}
        Z/Z_{\odot} = 10^{\rm{(O/H)} - \mathrm{(O/H)_{\odot}}}
    \end{equation}
    {Table~\ref{tab:metallicity} summarizes the metallicities (in solar units) for each source, measured across different regions selected based on their ionization states and using the methods described in Section~\ref{sec:metallicity_warm}. Specifically, since pAGB and X-ray/EUV photoionization models successfully reproduce the optical emission in most of our sources, we used these models to derive the O/H abundance in two regions: one near the center and one farther out. For sources dominated by LINER-like emission, we applied the corresponding LINER calibrations to the same two regions (e.g., Centaurus, M87). For specific regions ionized by HII, LINER, composite, we derived the O/H abundance applying the appropriate calibrations for each ionization mechanism.} 
    
    We used the same abundance table from \citet{Asplund2009} to calculate the hot gas metallicity, as detailed in Sec.~\ref{sec:hot_abundance}. The hot gas metallicity was calculated within regions similar to those used for the warm gas. We note that this was not possible for some sources due to low counts, so we computed the hot gas metallicity using a region that covered the entire warm gas. 
    
    To explore whether the warm and hot gas phase metallicities are similar, in Fig.~\ref{fig:warm_vs_hot_metallicity} we show the metallicity of the warm gas compared to that of the hot gas. We have also included O/H abundance measurements from a high-redshift cluster, MACS1931.8-2635 ($z =$ 0.3525), from MUSE observations discussed in \citealt{Ciocan21}. 
    
    As shown in Fig.~\ref{fig:warm_vs_hot_metallicity}, the warm and hot gas phase metallicities show a positive correlation. We used the {\sc linmix} Bayesian method \citep{kim17} to fit the warm -- hot gas phase metallicity correlation for the sources in our sample. One of the advantages of using Bayesian for linear regression is that the scatter is treated as a free parameter, together with the normalization, intercept and slope \citep{gaspari19}. The best-fit for the correlation is $\rm Z_{warm} = (0.15\pm0.09) + Z_{hot}^{0.90 \pm 0.10}$. Each parameter corresponds to the average of the distribution with 1$\sigma$ errors given by the standard deviation. The scatter correspond to 0.06$\pm$0.007, and the correlation to 0.75$\pm$0.03. The Pearson ($p$) value of the correlation is $p=0.77$. The correlation between the metallicity of both phases suggests that the gas phases are tightly coupled, and that the filaments form via turbulent condensation of the hot gas, as indicated by many theoretical/numerical studies \citep[e.g.,][]{Gaspari_2013,li15,li20,gaspari17,Voit_2017,beckmann19,Storchi-Bergmann_2019}.

    The scatter in the relation is likely driven by the different indirect methods used to measure the O/H abundance in the warm gas. In addition, the difference in metallicity between the hot and warm-gas phases can be attributed to the long mixing time, $\rm (t_{mix} = l/v_{l})$, where $l$ and $v_{l}$ are the length and turbulent velocity of the gas. For a typical cooling flow cluster, with a warm filament length of 10 kpc and velocity of 100 km/s, the turbulent mixing time is on the order of a few 100 Myr, comparable to the cooling time. Metals produced during star formation eventually mix with the hot gas phase after a few 100\,Myr, which explains the metallicity differences between the hot and warm gas phases. For chemical enrichment from Type Ia Supernovae, typical timescales range from $\sim$50 Myr for an instantaneous starburst to $\sim$0.3 Gyr for a typical elliptical galaxy \citep{Matteucci2001}.    
    
    Hydrodynamical simulations of self-regulated AGN feeding/feedback also predict a higher metallicity on the colder phase than the surrounding hot phase during the strong AGN feedback injection stage \citep{gaspari11a,gaspari11b}, as it entrains the central metals of the BCG into the surrounding atmosphere. After the AGN jets are triggered via chaotic cold accretion (CCA), extended filamentary cool gas starts to condense from the high-metallicity outflowing parcel of hot gas. As the AGN jets shut down (due to the quenched accretion), the cool gas will detach and rain back toward the center, where it will ultimately mix with the surrounding gas via turbulent diffusion, thus lowering the warm gas-phase metallicity. The AGN feedback could also spread centrally enriched metals to the outskirts of galaxies, into the halo and beyond on timescales of few 100 Myr \citep{beckmann19}. 
    Overall, the circulation flow driven by mechanical AGN feedback is a key mechanism to induce recursive variations in metal gradients, especially in massive/central galaxies.

    One of the implications of our work concerns the cold molecular gas content, as the CO-to-H2 conversion factor ($\alpha_{\rm CO}$) heavily relies on the metallicity \citep{Bolatto13}. All ALMA studies of massive cluster galaxies up to now \citep[e.g.,][]{salome11,russell19,olivares19} have assumed solar metallicity. We have found that the warm ionized gas, and likely the cold gas as well, shares a similar metallicity to the hot gas, which in most cases differs significantly from 1\,Z$_{\odot}$. The strong assumption on the metallicity can result on the understimation of the $\rm H_{2}$ mass traced by CO for some sources by almost an order of magnitude if the gas has a metallicity of 0.5~Z$_\odot$, when a galactic $\alpha_{\rm CO}$ is considered \citep{Amorin16}. The accurate measurement of the colder gas metallicity is crucial for addressing AGN feedback models, which report larger molecular gas masses than actual observations \citep[e.g.,][]{russell19,olivares19}.

\subsection{Star formation}

    Central cluster galaxies show a broad range of star formation rates (SFRs). Studies suggest that they are correlated with their intracluster medium properties as predicted by precipitation and chaotic cold accretion models \citep{gaspari11a,beckmann19}. Star formation, in turn, contributes to the ionization of the filaments as discussed by \cite{mcdonald11a}. As shown in this study, sources located in the composite region have larger SFRs, around between 10 -- 100 M$_{\odot}~yr^{-1}$ (e.g., Abell\,2597, Abell\,1835, PKS0745-19, Abell 1795), while sources dominated by LINER-like emission have very low SFRs, around 1 M$_{\odot}~yr^{-1}$ (Centaurus, M87, M84, and Abell\,496). This trend may be explained by the H$\alpha$ being enhanced by star formation processes, which move the spaxels towards the composite or HII-dominated regions. However,  measuring star formation rate solely from H$\alpha$ emission may lead to an overestimation of the SFR, due to the significant contribution of gas emission from composite or LINER-like regions.

\subsection{Limitations}\label{sec:limitations}

    A key limitation of this work concerns the O/H abundance measurements of the warm gas. Indirect methods, primarily calibrated for HII regions, vary significantly and typically yield higher metallicity values than direct methods based on electron temperatures $T_{e}$ \citep{Kewley08}—a pattern we observed in two of our systems. While the O/H abundance diagnostics we used have been calibrated for metallicities between 7.6 and 8.9, making them suitable for our systems, they were developed specifically for HII regions. This presents a challenge since our sample sources show various ionizing mechanisms, with most being composite in nature. We have tacked that by using different photoionizaiton models, such as pAGB stars and X-ray/EUV models, that can reproduce the emission line ratios of filaments.

    As discussed throughout this work, the $T_{\rm e}$-based method may provides a more accurate way to derive gas-phase metallicity \citep{Pilyugin2007}. Our Oxygen abundance measurements using calibration based on the ionization methods may give a higher abundance compared to the $T_{\rm e}$ method (see \citealt{Belfiore15} for more details). Future observations of the [OII]$\lambda\lambda$3726,3729 lines are essential, as MUSE cannot detect these in low-redshift sources needed to have more robust O/H estimates. Using instruments like ERIS, X-SHOOTER, or the upcoming BlueMUSE on the VLT would enable more precise warm gas abundance measurements through the ``direct'' Te method. Recent studies have also noted that the $T_{\rm e}$ method faces challenges due to large temperature fluctuations across galaxies, which can bias metallicity estimates lower than true metallicities \citep{Cameron2023}. In addition, $T_{\rm e}$ mode is considered more reliable for calculating  metallicity in low-metallicity environments \citep{pagel92,izotov06}, whereas in higher-metallicity environments, the $T_{\rm e}$ drops, making auroral lines more difficult to be detected. 
    
    It is worth noting that the metallicity measured in the hot gas through the $\sc{apec}$ modeling is driven by the Fe abundance, while the warm phase of the gas traces the Oxygen abundance. To date, studies using various different X-ray observatories have reported nearly solar abundance ratio of the hot gas at cluster centers, such as XMM-Newton \citep{mernier18}, Suzaku \citep{Sarkar2022}, and Hitomi \citep{hitomi17}, which indicate that both O and Fe abundances should be similar at the centers of clusters. A sizable uncertainty of approximately 20–30\% remains, due to instrumental systematics and inaccuracies in atomic codes. Ongoing and future XRISM observations, along with laboratory atomic experiments, will help establish more robust abundance ratios at cluster centers.

\section{Conclusions}
\label{sec:summary}
    In this paper we present metallicity measurements from both warm- and hot-gas phases in a sample of massive elliptical galaxies using high-resolution MUSE and \textit{Chandra} observations. 
    
\begin{enumerate}
    \item We found that warm gas in most of the massive elliptical galaxies in our sample is ionized by a combination of different mechanisms that lead to a composite state, with a clear contribution from star formation activity. Four sources — Centaurus, M87, M84, and Abell\,496 — which are the lowest redshift clusters, have the lowest star formation rate ($<$1\,M$_{\odot}\,yr^{-1}$) of the sample, and their ionized gas is dominated by LINER-like emission. The remaining sources in our sample — 2A0335, A1795, A1835, A2597, AS1101, Hydra-A, PKS0745 — have higher star formation rates ($>$1–200\,M$_{\odot}\,yr^{-1}$) and are dominated by composite ionization. Finally, only RXJ0821+0752 is dominated by HII ionization, with some contribution from LINER-like emission in the outer regions.

    \item We measured warm-gas metallicities using MUSE observations using different methods and derived hot-gas phase metallicities from \textit{Chandra} observations within regions where warm ionized gas is displayed. We found metallicities in both phases ranged between 0.3\,Z$_{\odot}$ and 1.5\,Z$_{\odot}$. 
    
    \item We found a positive linear correlation, $\rm Z_{warm} = (0.15\pm0.09) + Z_{hot}^{0.90 \pm 0.10}$, between the warm- and hot-gas phase metallicity, suggesting these phases are strongly connected. The scatter correspond to 0.06$\pm$0.007, and the correlation value to 0.75$\pm$0.03.
    The correlation suggests the warm gas originates from the thermally unstable cooling and condensation of the hot halo, which later triggers the AGN feedback. 

    \item The gas metallicity profiles exhibit a central drop in some systems, peaking at a few kiloparsecs before declining at larger radii. These patterns suggest that AGN mechanical feedback may have transported metals outward—as shown by hydrodynamical simulations—creating the central metallicity deficiency. Nevertheless, we cannot rule out the possibility that metals are partially depleted onto the colder gas phase, or that the observed metallicity abundance trends could arise from variations in the ionization conditions.
    
\end{enumerate}

Future observations from telescopes like X-SHOOTER and the upcoming blueMUSE on the VLT will provide more precise measurements of Oxygen abundance in the warm gas phase using the direct $T_{\rm e}$ method from [OII] and [OIII] emission lines—a capability MUSE lacks for most low-redshift clusters. Additionally, X-ray observations from the newly launched XRISM satellite will deliver more accurate measurements of hot gas metallicity and abundance for several sources discussed in this paper.

\begin{acknowledgements}
VO acknowledges support from the DICYT ESO-Chile Comite Mixto PS 1757, Fondecyt Regular 1251702.\\
PT acknowledges support from NASA’s NNH22ZDA001N Astrophysics Data and Analysis Program under award 24-ADAP24-0011.\\
MG acknowledges support from the ERC Consolidator Grant \textit{BlackHoleWeather} (101086804).\\
PL gratefully acknowledges support by the GEMINI ANID project No. 32240002.\\
ET acknowledges support from ANID programs CATA-BASAL FB210003 and FONDECYT Regular grants 1241005 and 1250821.\\

Based on observations collected at the European Organisation for Astronomical Research in the Southern Hemisphere under ESO programme(s): 094.A-0859(A), 095.B$-$0127(A), 60.A-9312(A), 0102.B-0048(A), 097.A-0366(A), 0103.A-0447(A), 095.B$-$0127(A), 097.A-0909(A) 
\end{acknowledgements}

\newpage
\begin{appendix}

\section{Summary of the observations}
        \begin{table*}[h!]
        \caption{Summary of the \textit{Chandra} and MUSE observation used in this paper.}
            \centering
            \begin{tabular}{p{2cm}cc|cc}
                \hline
        
                Source name & \multicolumn{2}{c}{Chandra observations} & \multicolumn{2}{c}{MUSE observations}\\
               \hline
               
                & {ID} & {exp. time} & {ID} & {exp. time}\\
                \hline
                2A0335+096 & 919, 7939, 9792 & 102.9\,ksec& 094.A-0859(A) & 2700s  \\ 
                
                Abell\,2597 & 922, 6934, 7329, 19596, 19597, 19598, 20626, & 625\,ksec &  094.A-0859(A) & 2700s  \\ 
                 & 20627, 20628, 20629, 20805, 20806, 20811, 20817 &  &  &   \\ 
                
                Abell\,S1101 & 11758, 1668 & 107.6\,ksec & 094.A-0859(A) & 2700s  \\ 
                
                Abell\,1795 & 493, 494, 3666, 5286, 5287, 5288, 5289,  & 257.5\,ksec & 094.A-0859(A) & 2700s  \\ 
                 &5290, 6159, 6160, 6161, 6162, 6163,17228  &  &  &   \\ 
        
                Abell\,1835 & 495, 496, 6880, 6881, 7370 & 223.9\,ksec & 097.A-0909(A) & 1177s  \\ 
                
                Abell\,496 & 4976, 931, 3361  & 104\,ksec &  095.B$-$0127(A) &  690$\times$4   \\ 
        
                Centaurus & 504, 4190, 4954, 4955, 5310, 16223 & \,ksec & 094.A-0859(A) \& 0103.A-0447(A) & 7271s \\ 
                 & 16224, 16225, 16534, 16607, 16608, 16609, 16610  &  &  &  \\ 
        
                Hydra$-$A & 575, 576, 4969, 4970 & 239.1\,ksec & 094.A-0859(A) &  2700s   \\ 
                
                M87 & 1808, 7210, 7211, 7212, 5826, 5827, 5828,  & 605.4\,ksec & 60.A-9312(A) & 5400s \\ 
                 & 18232, 18233, 18781, 18782, 18783 &  &  &  \\ 
        
                M84 &5908, 6131, 20539, 20540, 20541, 20542 & 853.7\,ksec & 0102.B-0048(A) & 2400s \\ 
                 & 20543, 21845, 21852, 21867, 22113, 22126, 22127 &  & &  \\ 
                & 22128, 22142, 22143, 22144, 22153, 22163, 22164,  &  & &  \\ 
                &  22166, 22174, 22175, 22176, 22177, 22195, 22196 &  & &  \\ 
        
                NGC5846 & 788, 7923 & 119\,ksec & 097.A-0366(A) & $7 × 880$ s  \\ 
                PKS0745$-$19 & 508, 2427, 6103, 12881 & 165.8\,ksec & 094.A-0859(A) &  2700s \\ 
                RXJ0821+0752 & 17194, 17563 &  272.9\,ksec & 094.A-0859(A) & 2700s  \\ 
        
                \hline        
            \end{tabular}
            \label{tab:observations}
        \end{table*}

\section{Example spectra}
\label{appendix:spectrum}

    \begin{figure}[h!]
    \centering
     \includegraphics[width=\columnwidth]{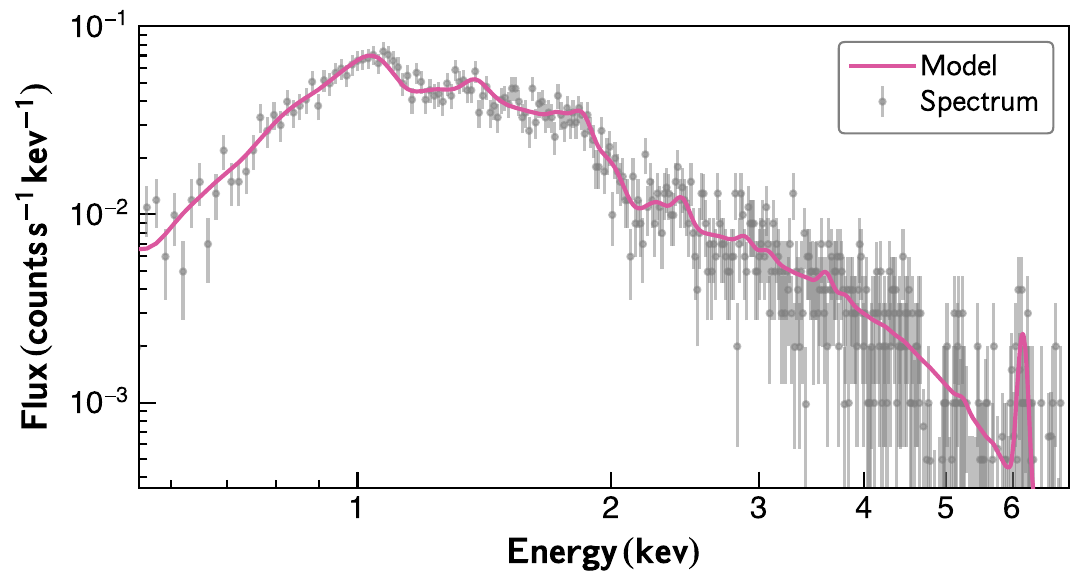}\\
     \includegraphics[width=\columnwidth]{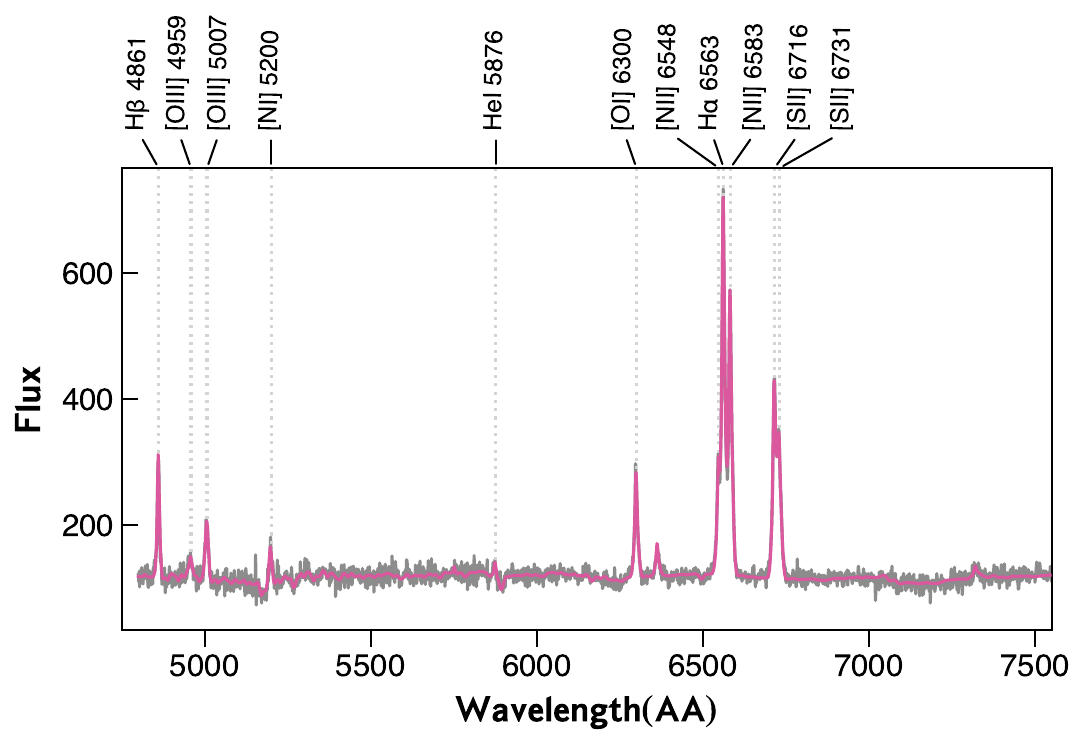}\\
     \caption{Top panel:Example X-ray spectrum of Abell\,2597 from one of the two regions used to derive the hot gas-phase metallicity. The observed spectrum, corresponding to one \textit{Chandra} observation, is shown as gray data points, while the best-fit model is overlaid in pink.
     Bottom Panel: Example of MUSE spectrum of one spaxel of Abell\,2597 shown in gray, and the best fitted model is shown in pink.}
     \label{fig:spectrum}
    \end{figure}

\section{Warm-phase abundances comparison}\label{appendix:warm_phase}
    \begin{figure}[h]
    \centering
        \includegraphics[width=0.9\columnwidth]{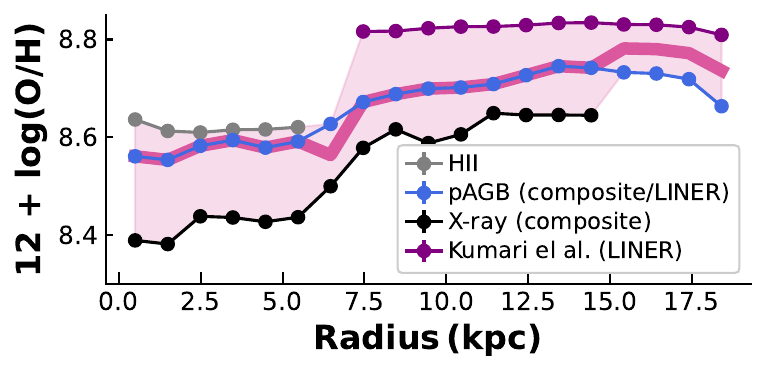}\\
        \caption{Comparison of O/H abundance profiles derived from the different O/H maps: X-ray/EUV and pAGB models, and HII and LINER calibration for RXJ0821. The median O/H profile is indicated by the pink line. O/H abundances profile derived from composite and LIN(E)R spaxels using X-ray and pAGB models, and LINER calibration are shown as blue, black, and purple points, respectively, while values obtained from H II spaxels are shown as gray points.}
     \label{appendix:fig_comparison_profile}
    \end{figure}

    \begin{figure}
    \centering
        \includegraphics[width=0.9\columnwidth]{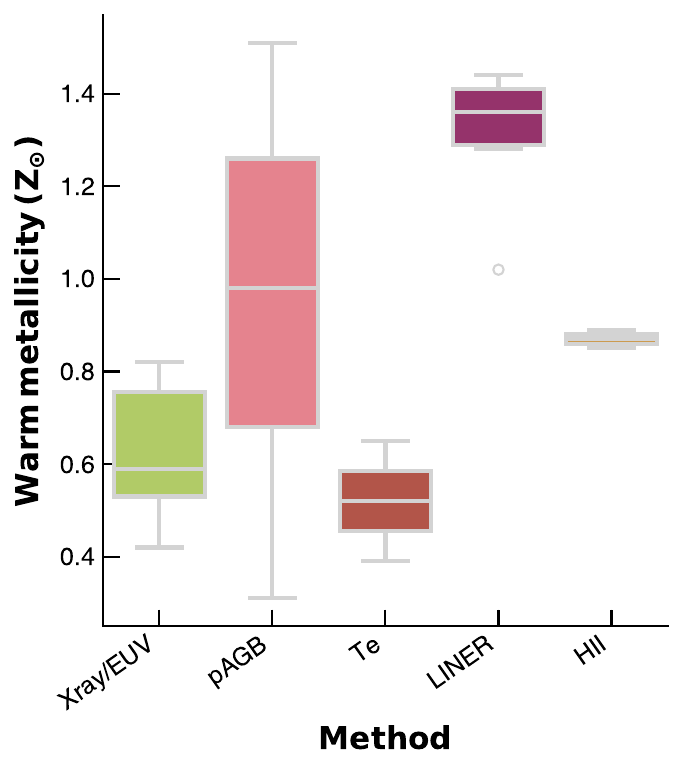}\\
        \caption{Diagram showing the metallicity ranges found in our sample using different methods throughout the paper. The green bar represents metallicity from X-ray/EUV models. Note that the X-ray/EUV method is limited to metallicities below 1\,Z$_{\odot}$. 
        The pink bar shows pAGB stars results, the red bar indicates $T_{\rm e}$ metallicities. The magenta bar shows LINER $O3N2$ measurements, while the gray bar correspond to results from HII $O3N2$ calibration.
     }
     \label{fig:methods}
    \end{figure}

    \begin{figure}[h!]
        \centering
        \includegraphics[width=0.7\columnwidth]{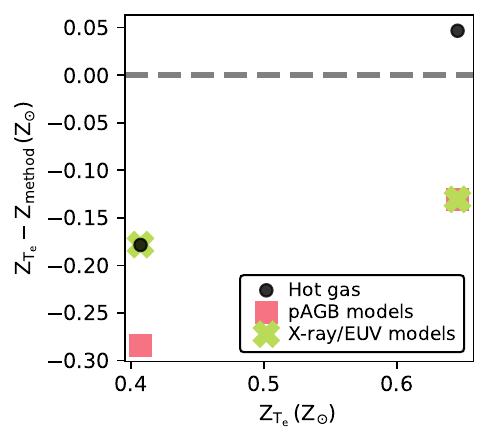}\\
         \caption{Comparison of the $T_{\rm e}$ abundance measurements with alternative methods. The plot show the the difference between metallicities obtained with $T_{\rm e}$ and other methods versus the $T_{\rm e}$-derived metallicity for two sources. Green represent metallicities derived from X-ray/EUV models, while pink squares correspond to values derived assuming photoionization by post-AGB stars. Black dots corresponds to the metallicity of the hot gas.}
     \label{fig:Te_comparison}
    \end{figure}

    In this section, we compared different methods for measuring warm-gas abundance. Figure \ref{appendix:fig_comparison_profile} shows the comparison of O/H abundance profiles derived from X-ray, pAGB models, and HII calibration for RXJ0821. The median O/H profile is indicated by the pink line. O/H abundances derived from composite and LIN(E)R spaxels using the X-ray and pAGB models are shown as blue and black points, respectively, while values obtained from HII spaxels are shown as gray points. On average, the pAGB model yields O/H abundances that are 0.15 dex higher than those from the X-ray model, although both models exhibit a consistent radial trend across the galaxy. This behavior, systematically higher pAGB abundances while following the same radial profile, is observed for all sources in our sample.
    
    In Figure~\ref{fig:methods}, we plot the warm-phase metallicity distribution for different methods to showcase their measurement ranges. This plot includes measurement from different regions of the gas. 
    
    The LINER/LIER calibration by \citet{Kumari2019} provided the highest O/H abundance values, with measurements ranging between 1.0 and 1.4\,$Z_{\odot}$, with the exception of the clusters that show clear AGN ionization sources at the center.
    
    For HII regions, the $O3N2$ calibration proposed by \citet{Marino2013} yielded O/H abundances between 0.85 and 0.89\,$Z_{\odot}$.
    The X-ray/EUV models showed abundances ranging from 0.4 to 0.8\,$Z_{\odot}$. However, these models cannot account for metallicities above 1.0\,$Z_{\odot}$, making it impossible to measure several high-metallicity systems such as Centaurus, M87, M84, and A496. The emission line ratios of these sources consistently indicate metallicities $>1Z_{\odot}$.
    
    The pAGB models yielded O/H abundances between 0.3 and 1.5$Z_{\odot}$, similar to the X-ray/EUV models but systematically higher by a median of 0.20\,$Z_{\odot}$. 

    Using the T$_{\rm e}$ method, we found relatively low O/H abundances of 0.5--0.3\,Z$_{\odot}$ for two sources that have [OII] emission lines. In Fig.~\ref{fig:Te_comparison}, we compare the differences between X-ray/EUV measurements, pAGB models. For consistency, we computed the abundance using the integrated values the two sources following the methods described in the Sec.~\ref{sec:metallicity_warm}. The metallicity derived from T$_{\rm e}$ measurements shows the closest agreement with X-ray/EUV derived metallicities, having differences of 0.1–0.17\,Z${\odot}$, and the pAGB models with 0.13–0.18\,Z$_{\odot}$.

\section{BPT diagrams, and O/H abundance maps and profiles }
\label{appendix:BPT}

\setlength\fboxrule{0.5mm}   
\setlength\fboxsep{0.5pt}       

    \begin{figure*}
        \fcolorbox{gray}{white!5}{
        \parbox[b]{\textwidth}{
        \centering
          \includegraphics[width=0.52\columnwidth]{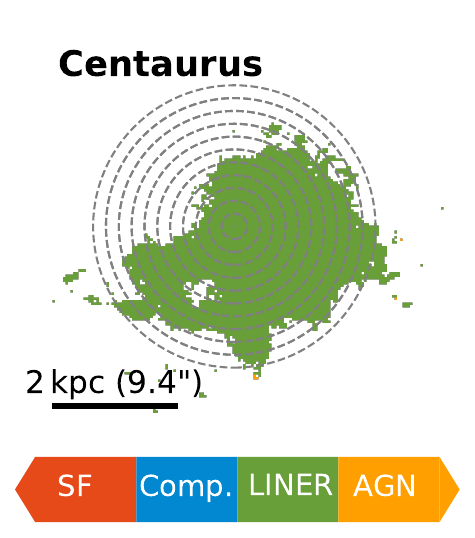}
          \includegraphics[width=0.52\columnwidth]{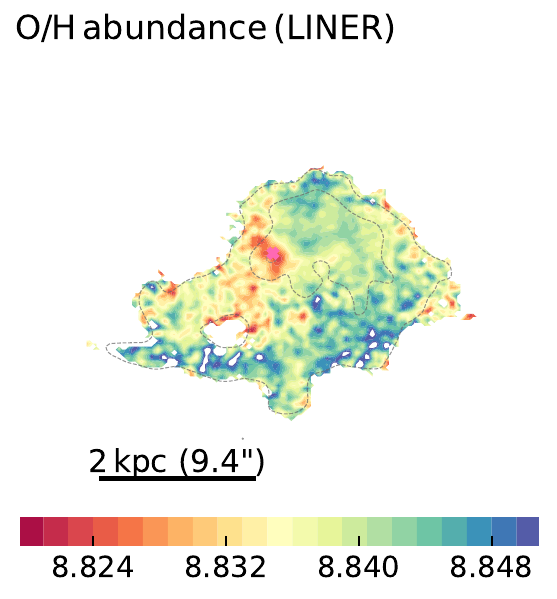}
        \includegraphics[width=0.9\columnwidth]{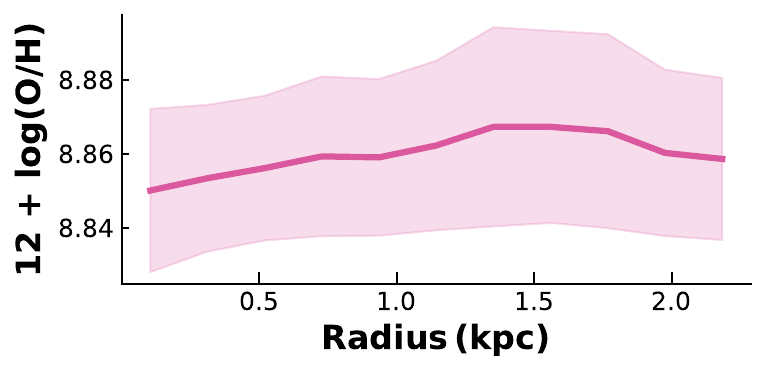}\\
    \hspace{-13cm}\includegraphics[width=0.52\columnwidth]{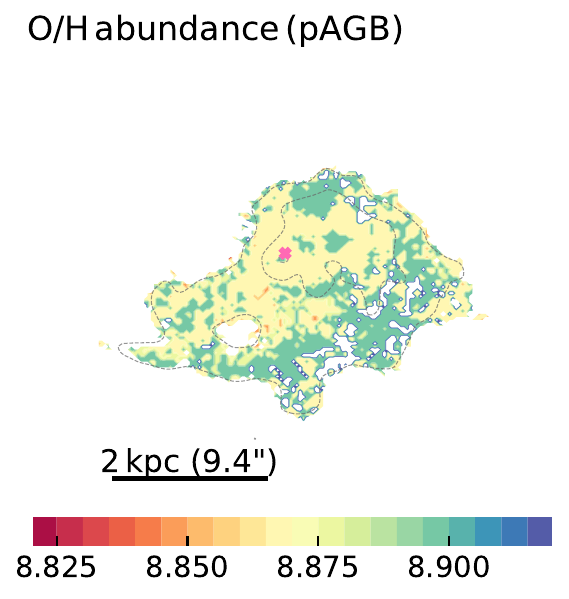}\\

        }}\\

        \fcolorbox{Gray}{White!5}{
        \parbox[b]{\textwidth}{
        \includegraphics[width=0.52\columnwidth]{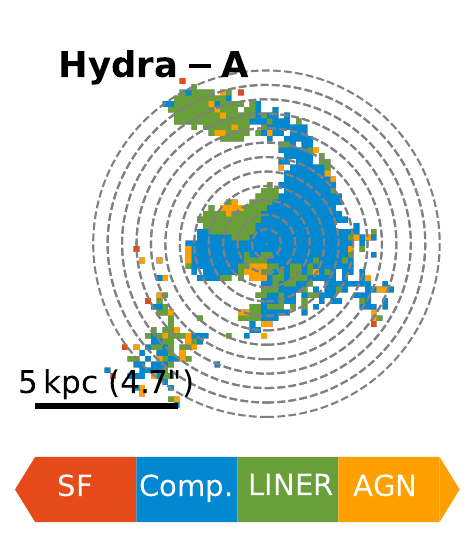}
          \includegraphics[width=0.52\columnwidth]{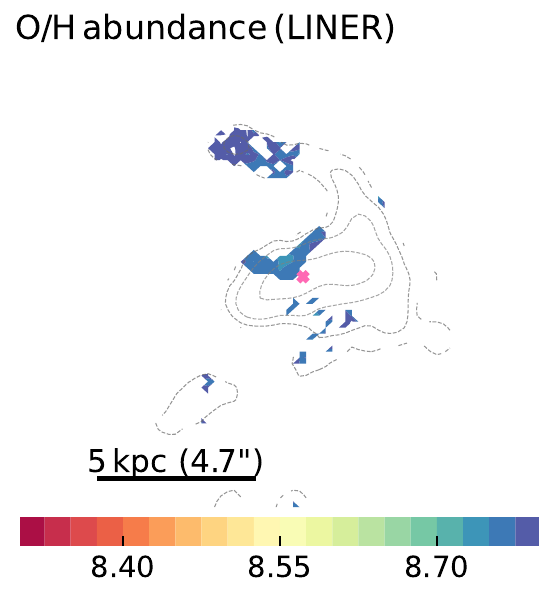}
        \includegraphics[width=0.9\columnwidth]{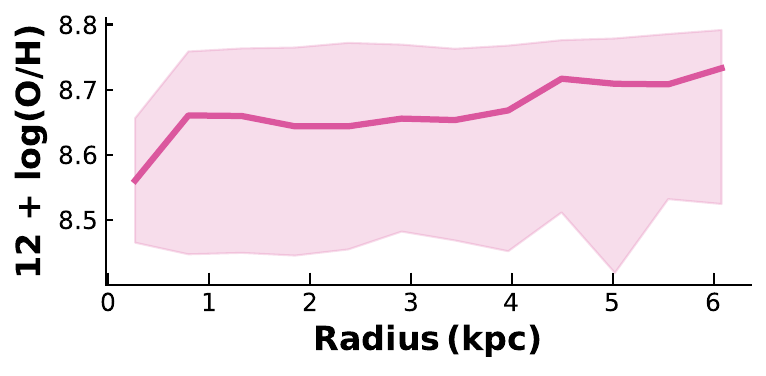}\\
        \hspace{3cm}\includegraphics[width=0.52\columnwidth]{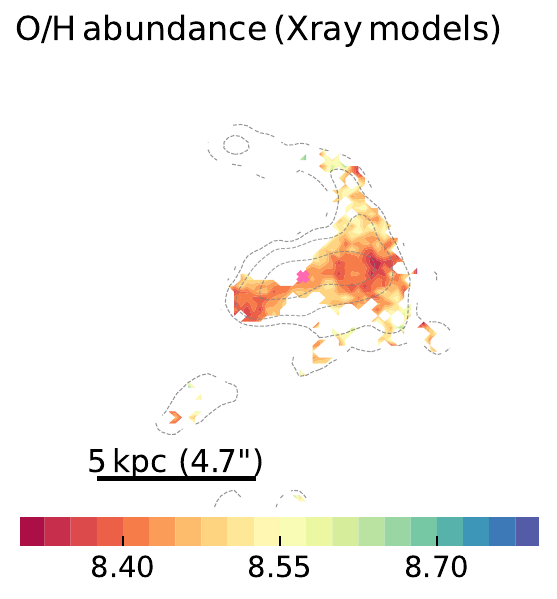}
        \includegraphics[width=0.52\columnwidth]{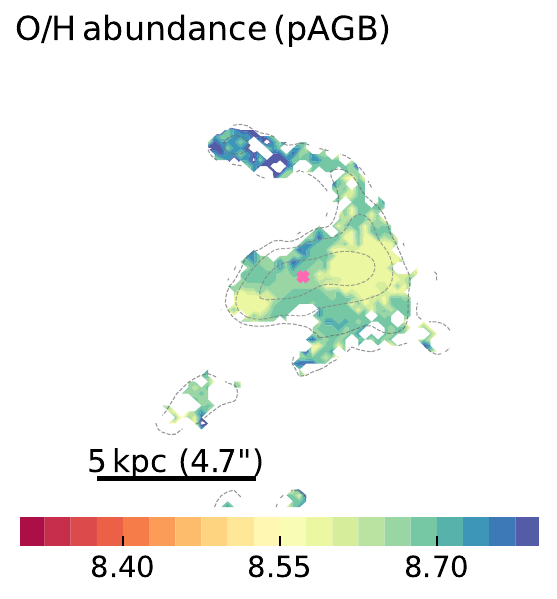}

        }}
        \caption{Examples of O/H abundance profiles from our sample, illustrating both cases where central O/H drops are detected (e.g., Centaurus, RXJ0821+0752).
        Upper left panel: Spatially resolved BPT diagram showing regions classified as AGN (yellow), LI(N)ER (green), composite (blue), and HII (red). The gray circles indicate the annulus used to derive O/H abundances.
        Upper middle panel: O/H abundance map derived using the LI(N)ER calibration for the LI(N)ER spaxels.
        Upper right panel: Median radial profile of the warm gas O/H abundance, computed using calibrations and models appropriate for each spaxel depending on the ionization mechanism. 
        The shaded region shows the 85\% confidence interval, estimated via bootstrapping, and reflects the range of O/H values obtained from the different methods.
        Bottom left, middle, and right panels: O/H abundance maps derived using the X-ray/EUV and pAGB models for the composite and LINER ionization, respectively. O/H abundance maps derived using HII calibration for the SF dominated spaxels (if applies). The pink cross shows the center of the galaxy.}\label{fig:abundance_maps}
    \end{figure*}

    \begin{figure*}
        \fcolorbox{Gray}{White!5}{
        \parbox[b]{\textwidth}{
        \includegraphics[width=0.52\columnwidth]{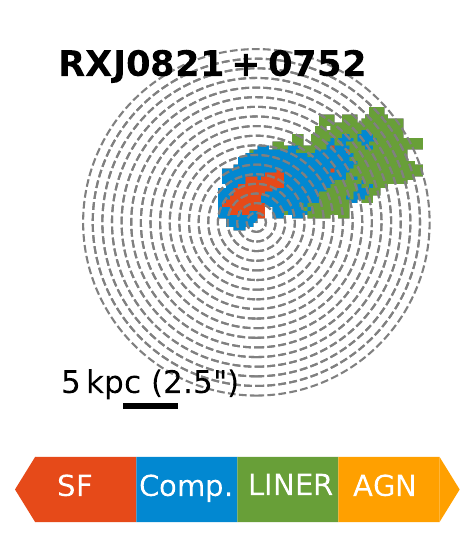}
        \centering \includegraphics[width=0.52\columnwidth]{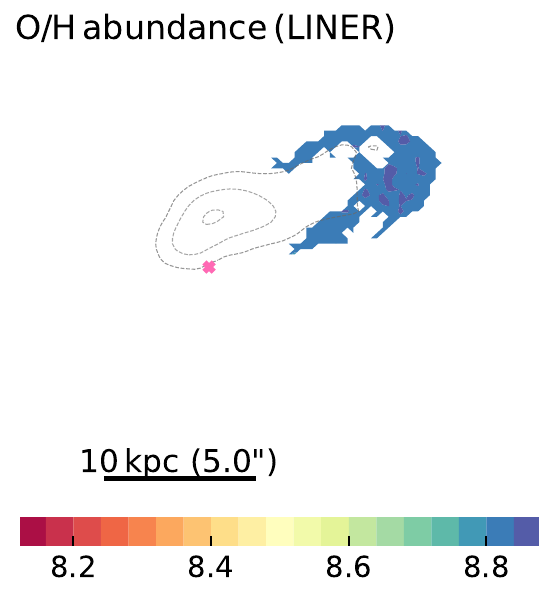}
         \includegraphics[width=0.9\columnwidth]{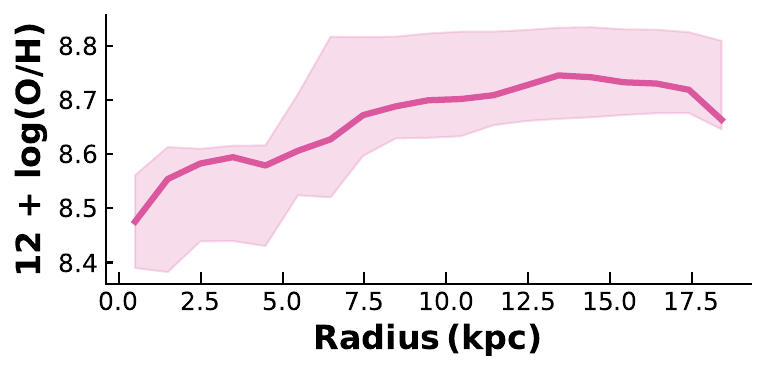}\\
        \hspace{-3cm}\includegraphics[width=0.52\columnwidth]{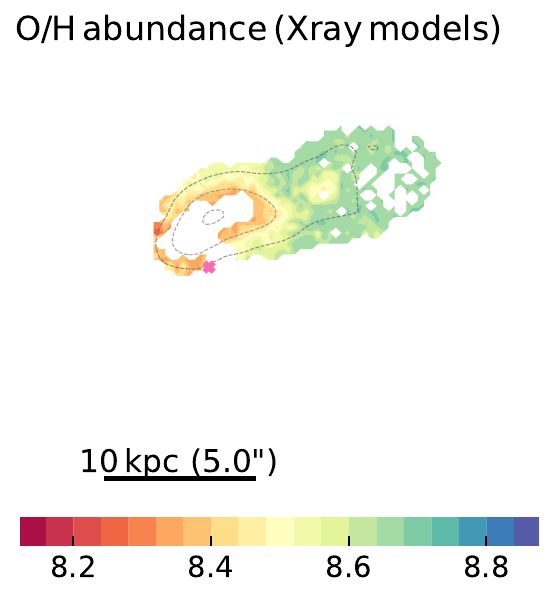}
        \includegraphics[width=0.52\columnwidth]{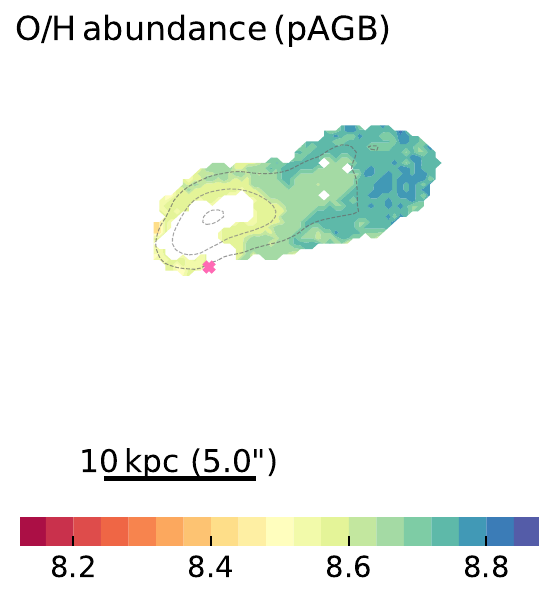}
         \includegraphics[width=0.52\columnwidth]{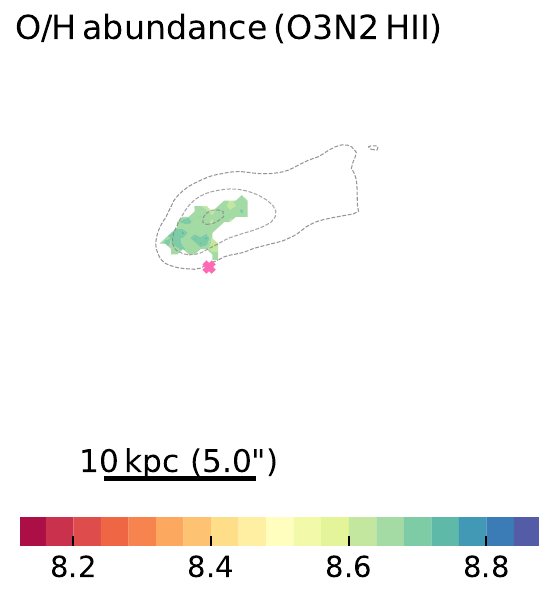}
        }}\\

        \fcolorbox{Gray}{White!5}{
        \parbox[b]{\textwidth}{
        \centering
          \includegraphics[width=0.52\columnwidth]{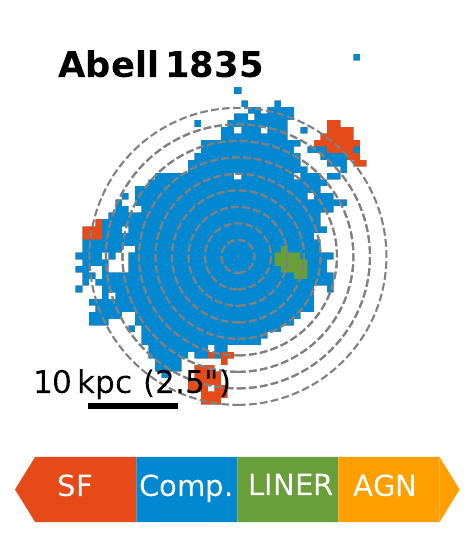}
          \includegraphics[width=0.52\columnwidth]{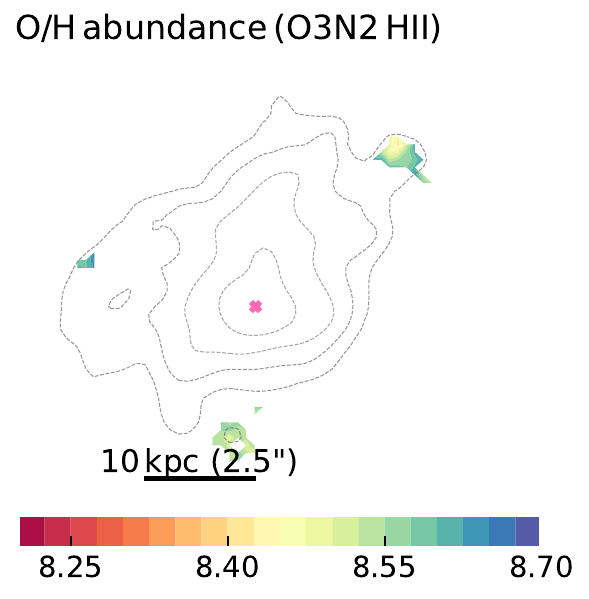}
        \includegraphics[width=0.9\columnwidth]{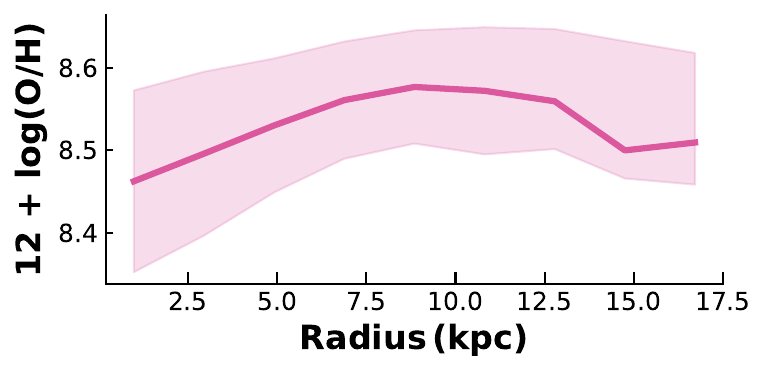}\\
        \hspace{-8cm}\includegraphics[width=0.52\columnwidth]{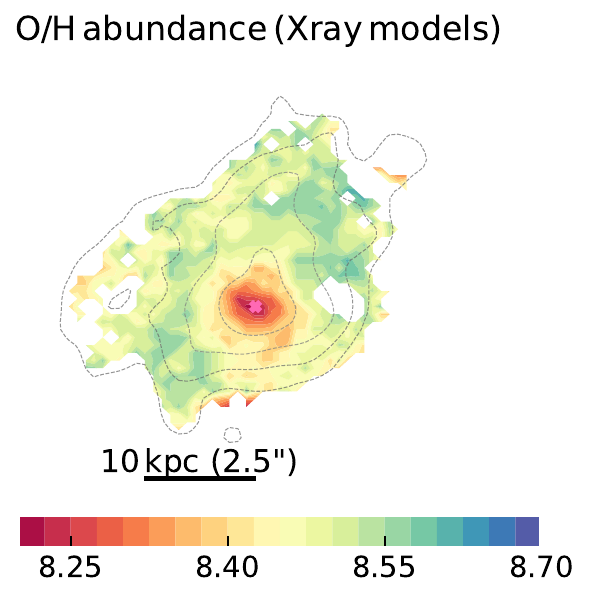}
        \includegraphics[width=0.52\columnwidth]{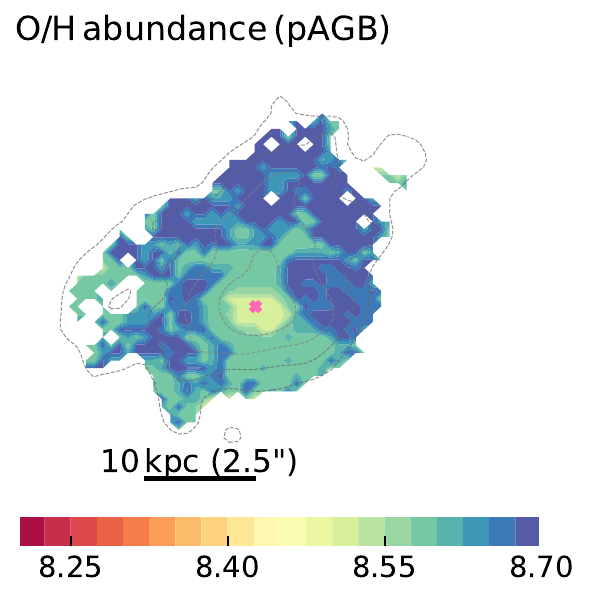}

        }}
             \caption{Same as Fig.~\ref{fig:abundance_maps}}
         \label{fig:abundance_maps2}
    \end{figure*}

\begin{figure*}[h!]
\centering
\fcolorbox{Gray}{White!5}{
\parbox[b]{\textwidth}{
  \includegraphics[width=0.52\columnwidth]{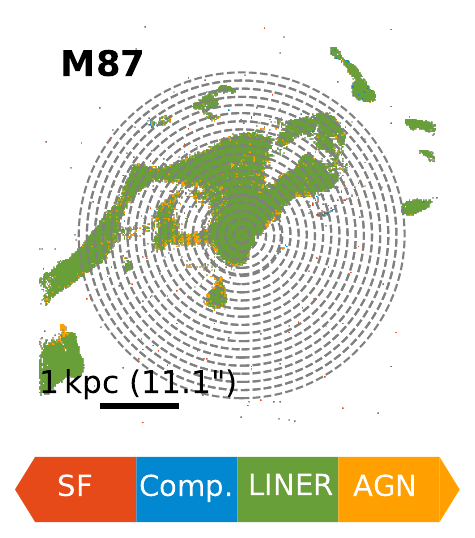}
  \includegraphics[width=0.52\columnwidth]{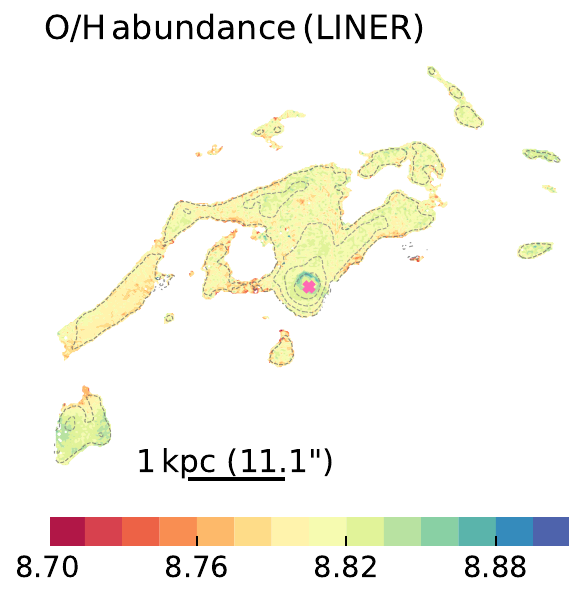}
\includegraphics[width=0.9\columnwidth]{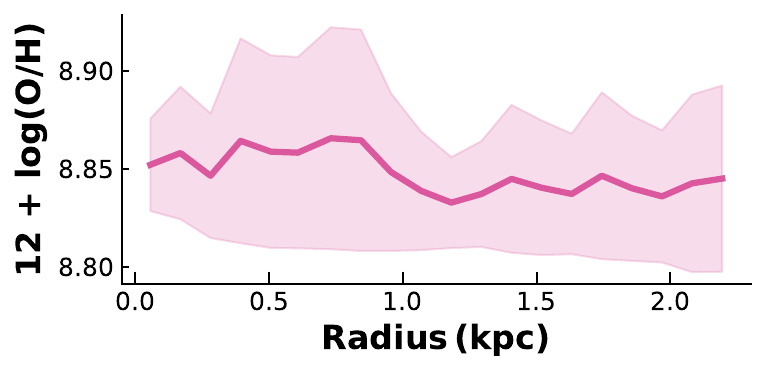}\\
  \includegraphics[width=0.52\columnwidth]{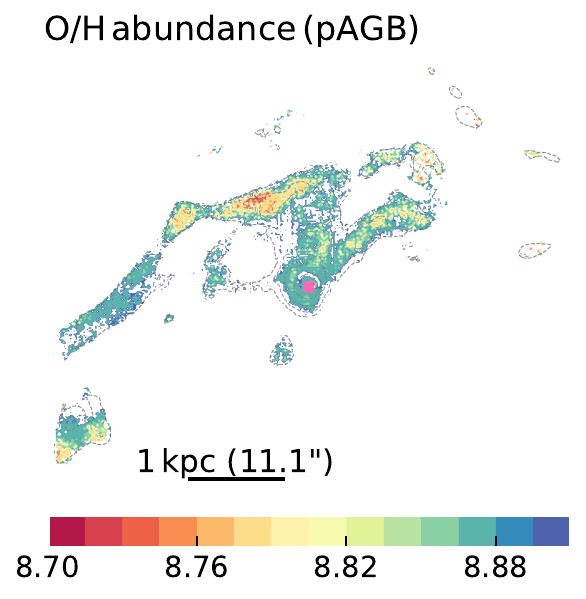}

}}\\

    \fcolorbox{Gray}{White!5}{
    \parbox[b]{\textwidth}{
    \includegraphics[width=0.52\columnwidth]{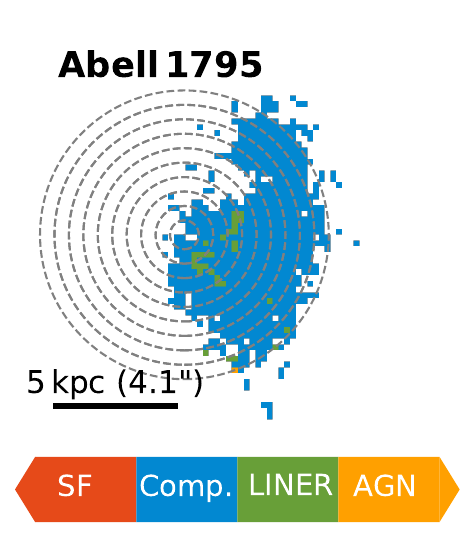}
      \hspace{4cm}
        \includegraphics[width=0.9\columnwidth]
        {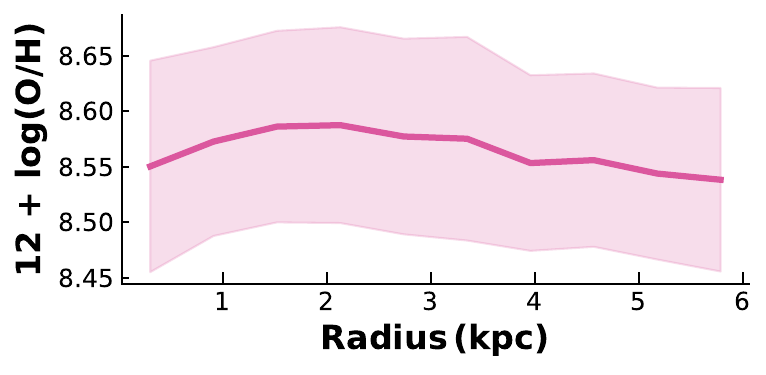}\\
        
        \includegraphics[width=0.52\columnwidth]{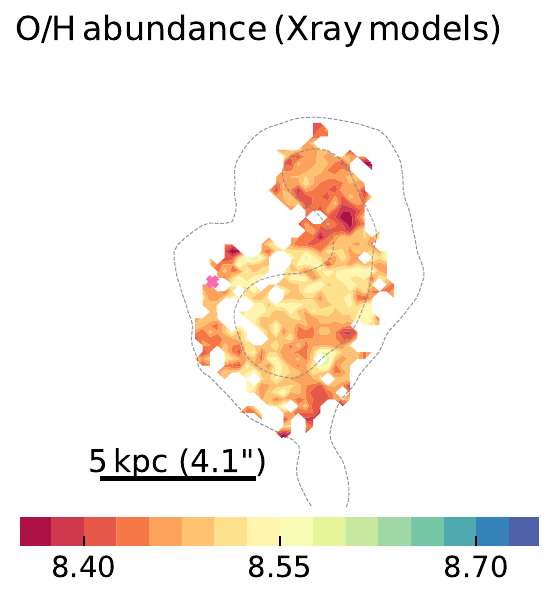}
    \includegraphics[width=0.52\columnwidth]{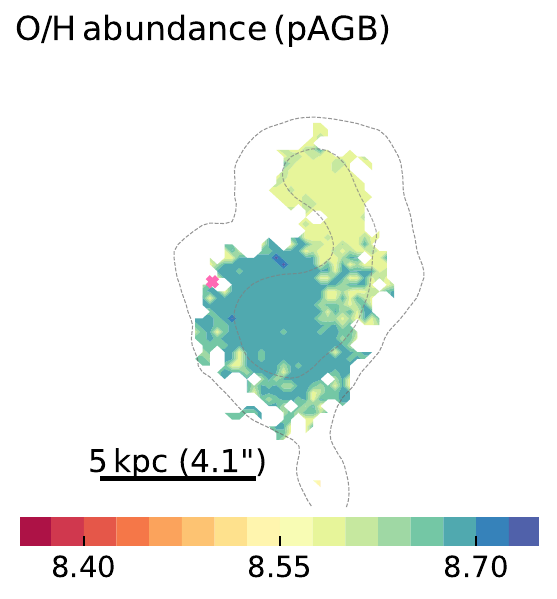}

    }}\\

 \caption{Same as Fig.~\ref{fig:abundance_maps}.}
 \label{fig:app_abundance_maps1}
\end{figure*}

\begin{figure*}[h!]
    \fcolorbox{Gray}{White!5}{
    \parbox[b]{\textwidth}{
      \includegraphics[width=0.52\columnwidth]{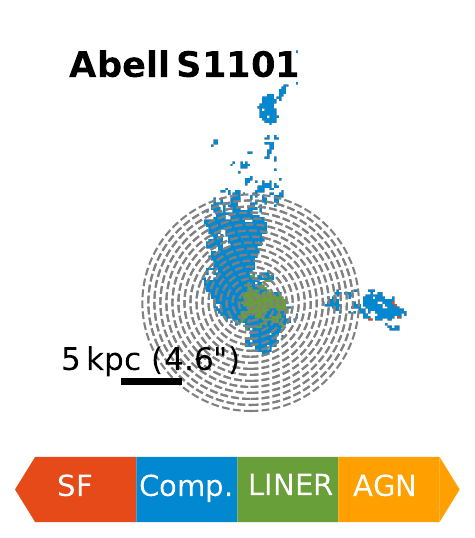}
      \includegraphics[width=0.52\columnwidth]{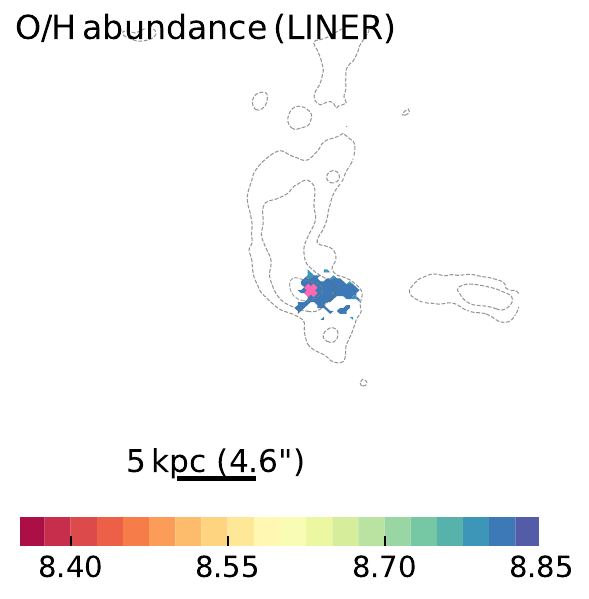}
        \includegraphics[width=0.9\columnwidth]{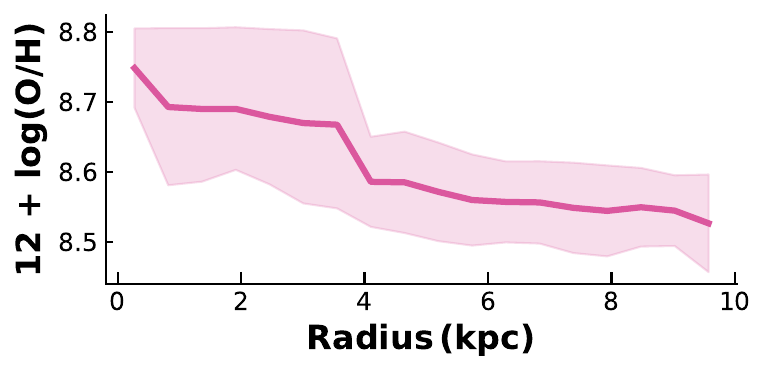}\\
      \includegraphics[width=0.52\columnwidth]{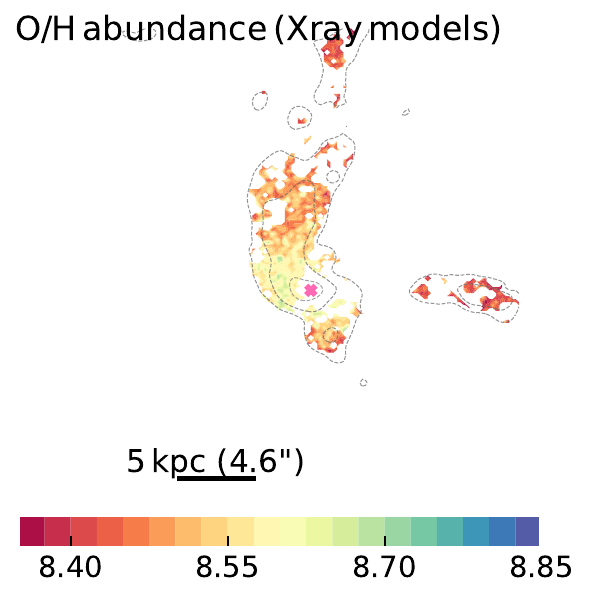}
        \includegraphics[width=0.52\columnwidth]{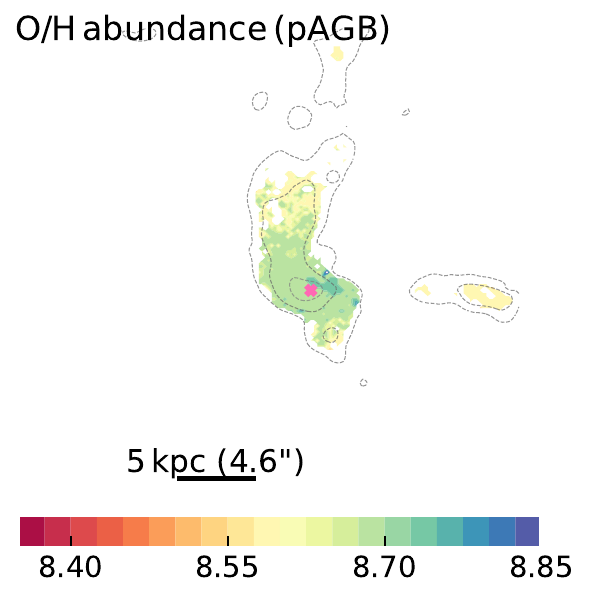}
    }}

    \fcolorbox{Gray}{White!5}{
    \parbox[b]{\textwidth}{
      \includegraphics[width=0.52\columnwidth]
      {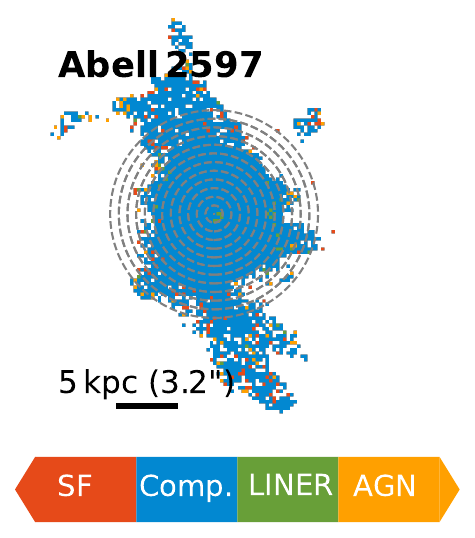}
      \hspace{4cm}
        \includegraphics[width=0.9\columnwidth]{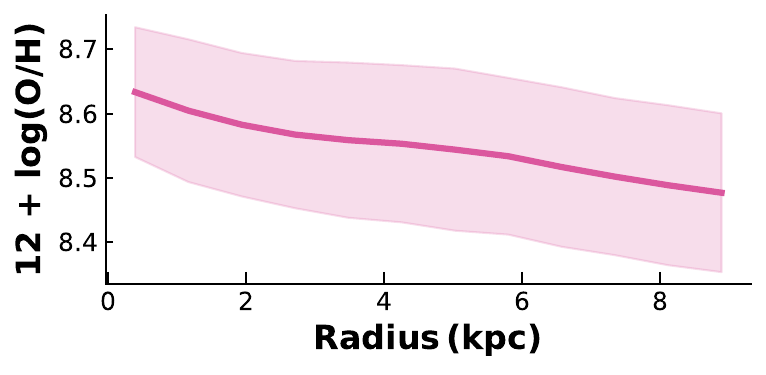}\\
      \includegraphics[width=0.52\columnwidth]{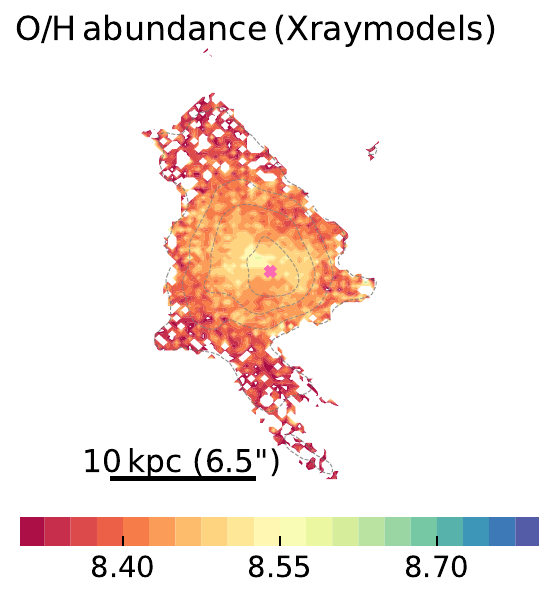}
        \includegraphics[width=0.52\columnwidth]{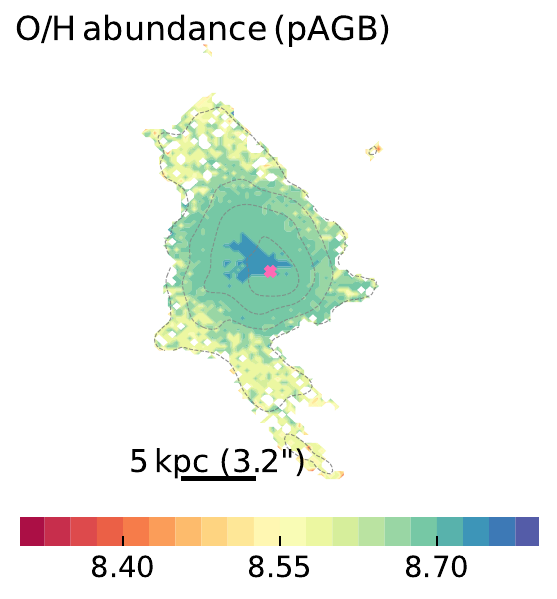}

    }}
    
    \caption{Same as Fig.~\ref{fig:abundance_maps}.}\label{fig:app_abundance_maps2}
    \end{figure*}
    
\begin{figure*}[h!]
    \fcolorbox{Gray}{White!5}{
    \parbox[b]{\textwidth}{
      \includegraphics[width=0.52\columnwidth]{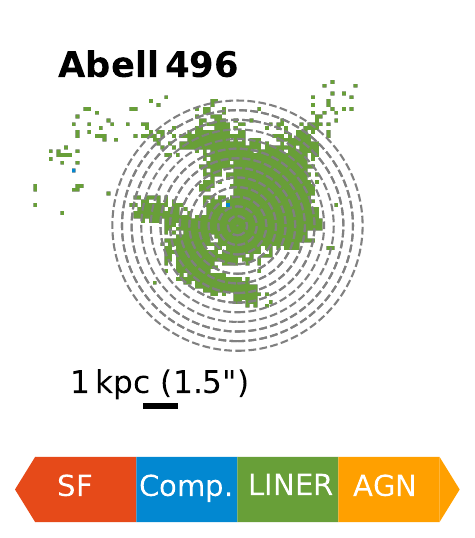}
      \includegraphics[width=0.52\columnwidth]{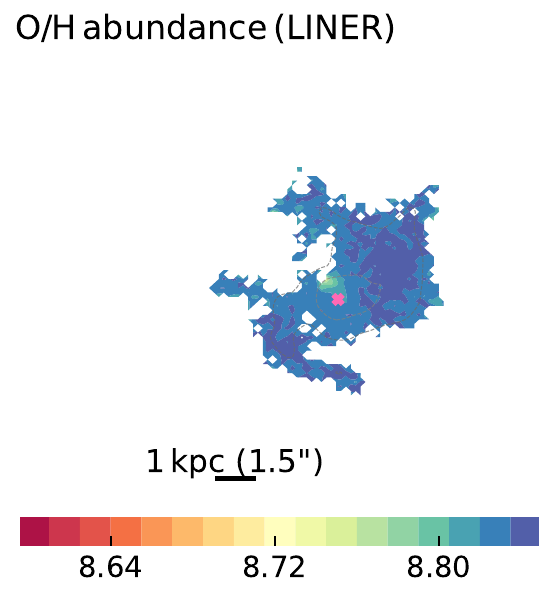}
        \includegraphics[width=0.9\columnwidth]{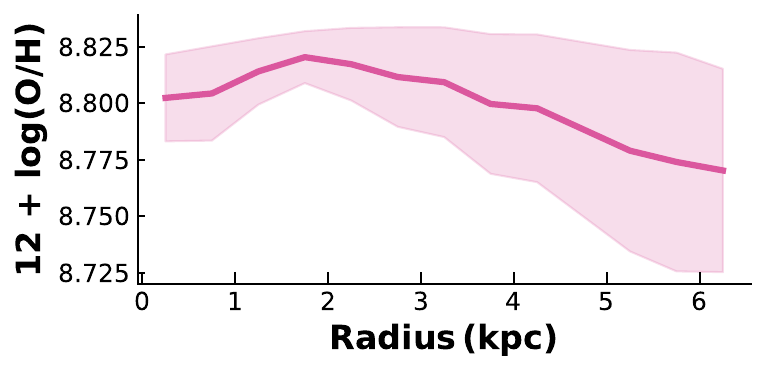}\\
        \includegraphics[width=0.52\columnwidth]{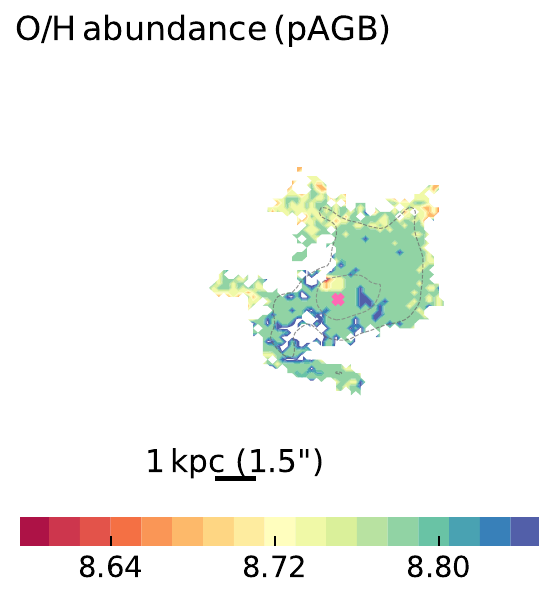}
    }}\\

    \fcolorbox{Gray}{White!5}{
    \parbox[b]{\textwidth}{
      \includegraphics[width=0.52\columnwidth]{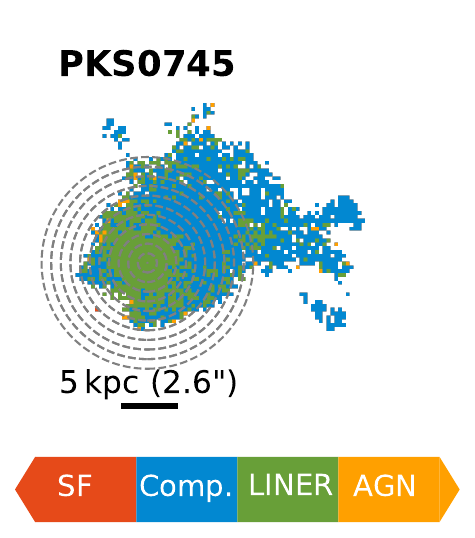}
      \includegraphics[width=0.52\columnwidth]{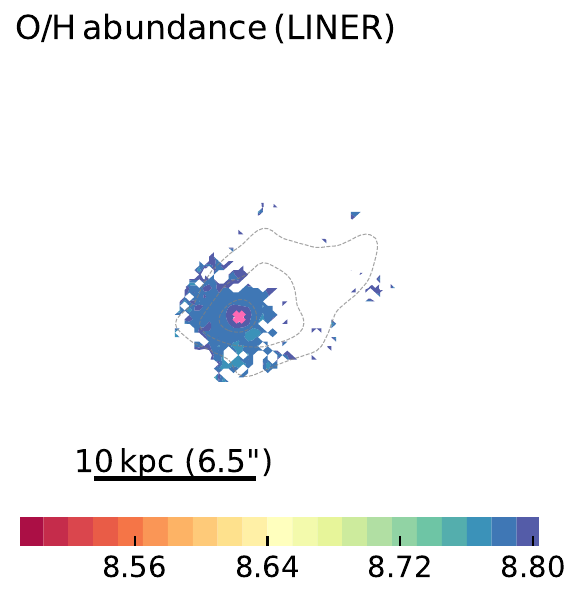}
        \includegraphics[width=0.9\columnwidth]{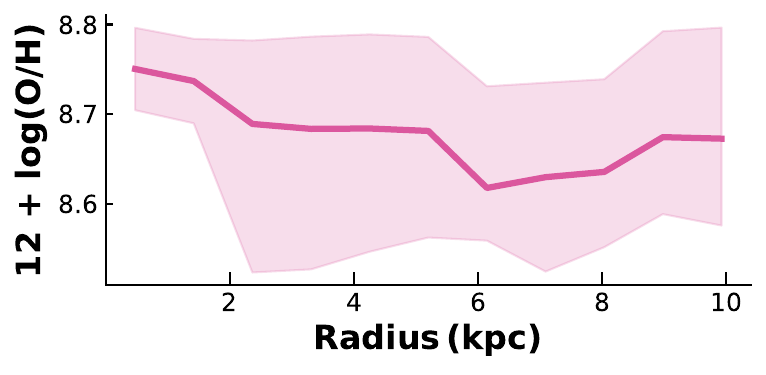}\\
    \includegraphics[width=0.58\columnwidth]{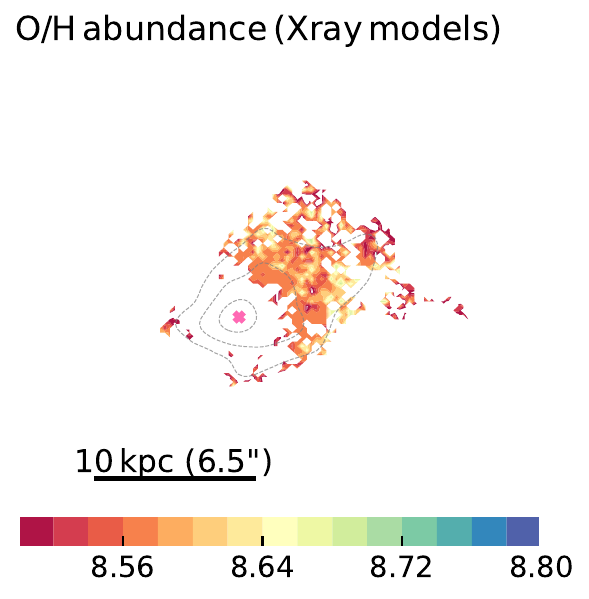}
    \includegraphics[width=0.58\columnwidth]{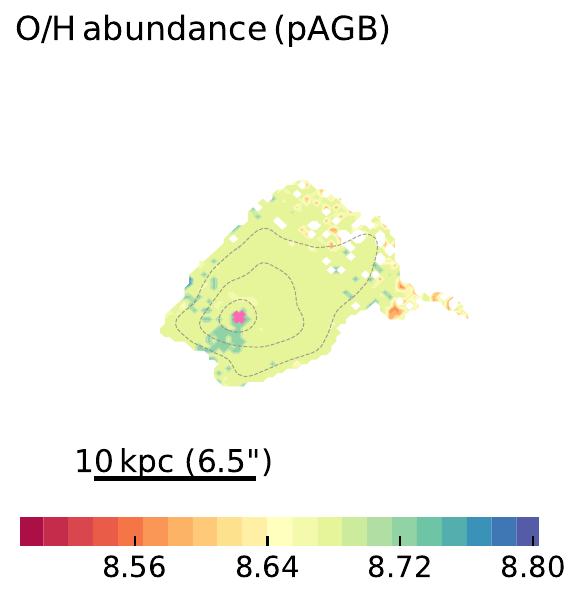}
    }}\\
    
 \caption{Same as Fig.~\ref{fig:abundance_maps}.}
 \label{fig:app_abundance_maps3}
\end{figure*}

\begin{figure*}[h!]
    \fcolorbox{Gray}{White!5}{
    \parbox[b]{\textwidth}{
      \includegraphics[width=0.52\columnwidth]{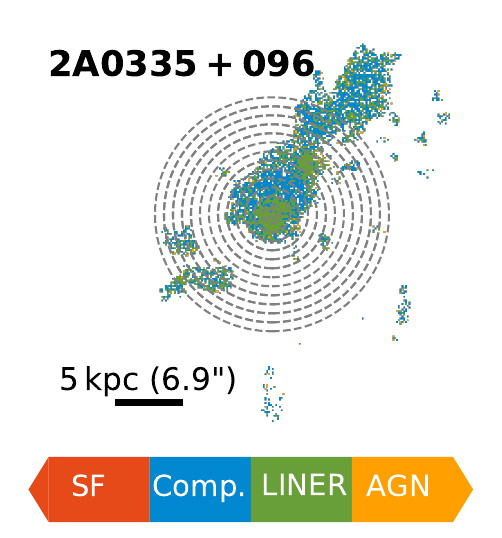}
      \includegraphics[width=0.52\columnwidth]{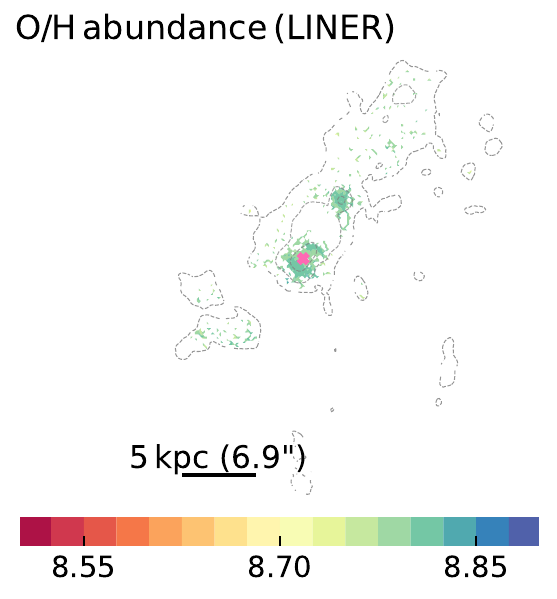}
        \includegraphics[width=0.9\columnwidth]{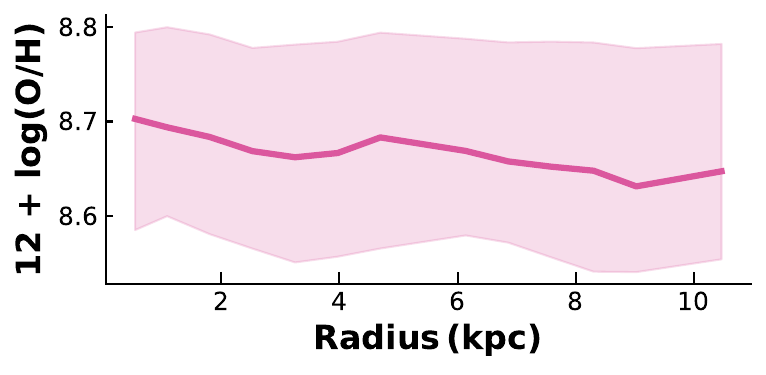}\\
        \includegraphics[width=0.52\columnwidth]{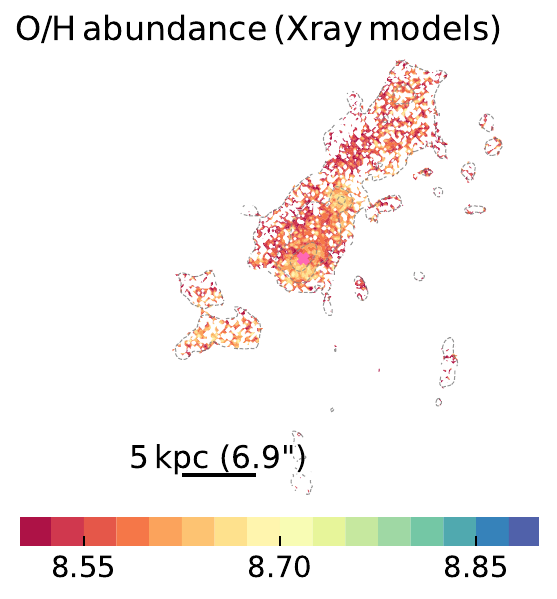}
        \includegraphics[width=0.52\columnwidth]{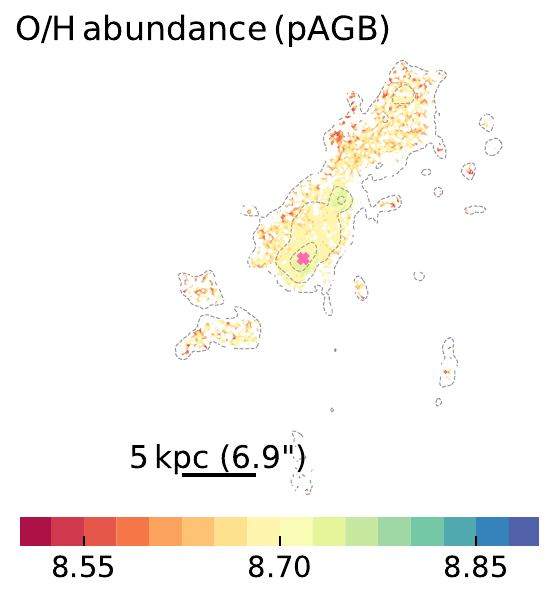}
    }}\\

    \fcolorbox{Gray}{White!5}{
    \parbox[b]{\textwidth}{
      \includegraphics[width=0.52\columnwidth]{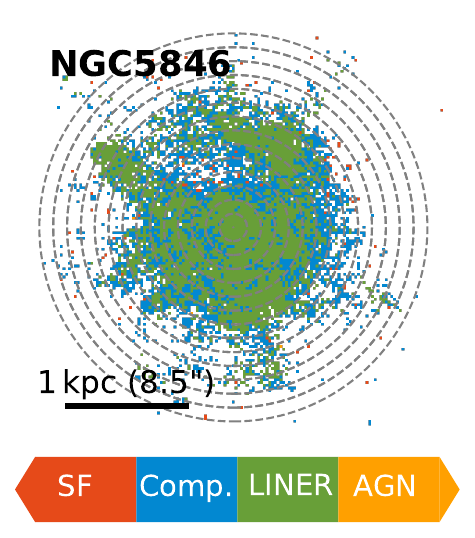}
      \includegraphics[width=0.52\columnwidth]{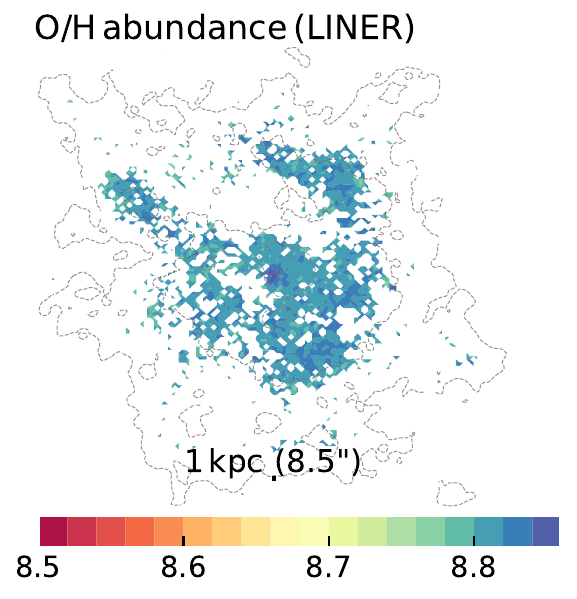}
        \includegraphics[width=0.9\columnwidth]{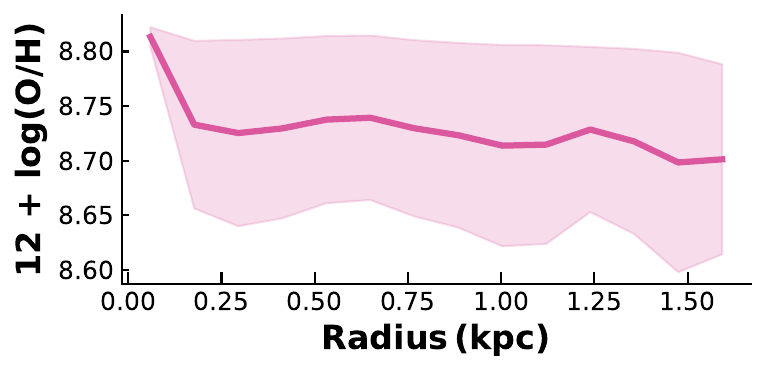}\\
        \includegraphics[width=0.52\columnwidth]{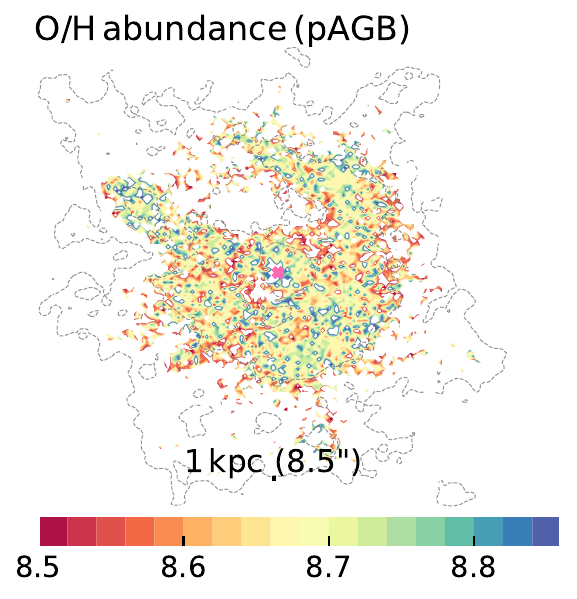}
    }}

 \caption{Same as Fig.~\ref{fig:abundance_maps}.}
 \label{fig:app_abundance_maps4}
\end{figure*}

\begin{figure*}[h!]
    \fcolorbox{Gray}{White!5}{
    \parbox[b]{\textwidth}{
      \includegraphics[width=0.52\columnwidth]{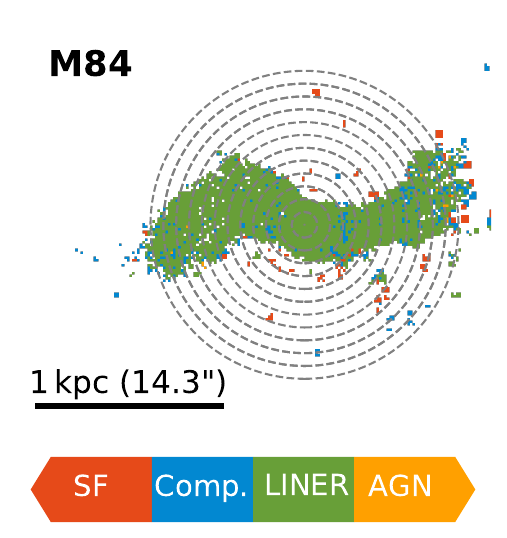}
      \includegraphics[width=0.52\columnwidth]{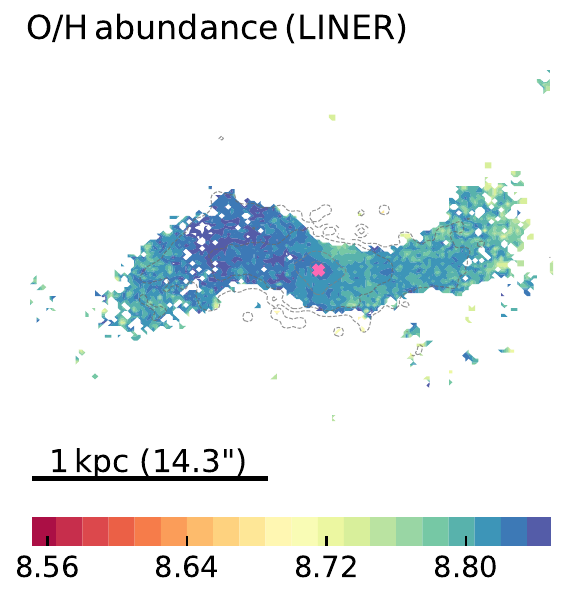}
        \includegraphics[width=0.9\columnwidth]{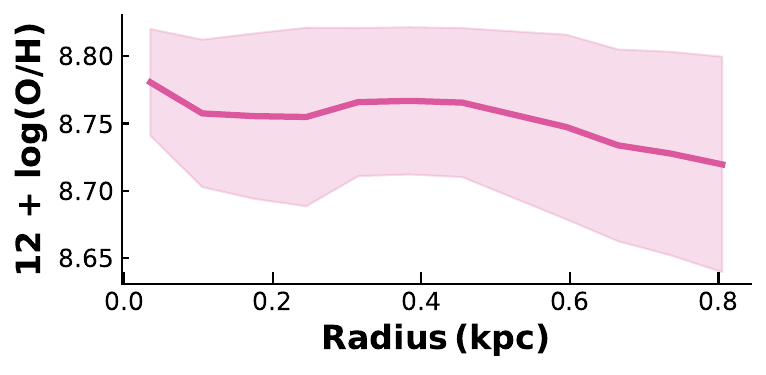}\\
        \includegraphics[width=0.52\columnwidth]{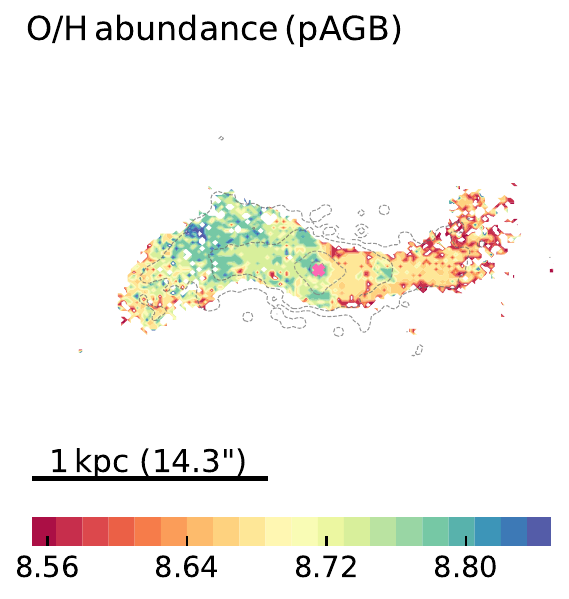}
    }}\\

 \caption{Same as Fig.~\ref{fig:abundance_maps}.}
 \label{fig:app_abundance_maps5}
\end{figure*}

\end{appendix}

\end{document}